\documentclass[aps,twocolumn,superscriptaddress,notitlepage,longbibliography]{revtex4-1}%
\usepackage{float}
\usepackage{graphicx}
\usepackage{amsmath,amssymb,bbm,mathrsfs,bm,braket,color,graphicx,comment,amsfonts,dsfont}
\usepackage[colorlinks,citecolor=blue,urlcolor=blue]{hyperref}
\usepackage[mathscr]{euscript}
\usepackage[normalem]{ulem}
\usepackage{comment}

\DeclareMathAlphabet\mathbfcal{OMS}{cmsy}{b}{n}

\usepackage{extarrows} 
\usepackage{mathrsfs}

\definecolor{purple}{rgb}{0.3, 0.8, 0.}
\definecolor{grun}{rgb}{0.0, 0.7, 0.0}

\usepackage{tikz,lipsum,lmodern}
\usepackage[most]{tcolorbox}

\begin{document}

\title{Fate of surface gaps in magnetic topological insulators}

\author{Habib Rostami}\email{ hr745@bath.ac.uk}
\address{Department of Physics, University of Bath, Claverton Down, Bath BA2 7AY, United Kingdom}

\author{Ali G. Moghaddam}\email{agorbanz@iasbs.ac.ir}
\address{Computational Physics Laboratory, Physics Unit, Faculty of Engineering and Natural Sciences, Tampere University, P.O. Box 692, FI-33014 Tampere, Finland}
\address{Helsinki Institute of Physics, FI-00014 University of Helsinki, Finland}

\date{\today}

\begin{abstract}
In magnetic topological insulators, the surface states can exhibit a gap due to the breaking of time-reversal symmetry. Various experiments, while suggesting the existence of the surface gap, have raised questions about its underlying mechanism in the presence of different magnetic orderings. Here, we demonstrate that magnon-mediated electron-electron interactions, whose effects are not limited to the surfaces perpendicular to the magnetic ordering, can significantly influence surface states and their effective gaps. On the surfaces perpendicular to the spin quantization axis, many-body interactions can enhance the band gap to a degree that surpasses the non-interacting scenario. Then, on surfaces parallel to the magnetic ordering, we find that strong magnon-induced fermionic interactions can lead to features resembling a massless-like gap. 
These remarkable results largely stem from the fact that magnon-mediated interactions exhibit considerable long-range behavior compared to direct Coulomb interactions among electrons, thereby dominating the many-body properties at the surface of magnetic topological insulators.
\end{abstract}

\maketitle

\section*{Introduction}
The commencement of the new century has coincided with groundbreaking discoveries in condensed matter physics. Notably, the theoretical prediction and experimental realization of various topological phases of matter have revolutionized our understanding of even the most fundamental electronic properties of solids \cite{Kane2010RMP,Zhang2011RMP,ando2013topological}. Beyond their fundamental significance, topological materials have proven to be highly promising for advanced technological applications such as nanoelectronics, spintronics, and quantum technologies \cite{ando2022opportunities,Ozawa2019,Nayak2008,luo2022topological}. A quintessential hallmark of three-dimensional topological phases, upon which much of the intriguing physics and potential applications hinge, is the existence of two-dimensional (2D) surface states that offer protection against perturbations, disorder, and imperfections \cite{moore2010birth,Hatsugai,graf2013bulk,prodan2016bulk}. 

Recently, the discovery of magnetic topological insulators (MTIs), wherein nontrivial topology intersects with magnetic ordering in either antiferromagnetic (AFM) or ferromagnetic (FM) forms, has introduced a new dimension to topological physics \cite{bernevig2022progress,Tokura2019,katmis2016high,Manchon2021,deng2020quantum,wang2021intrinsic,Recent_MnBi2nTe3n1,wang2023materials}. From a practical perspective, these materials hold promise for applications in spintronics, magnetic memories, and other domains \cite{vsmejkal2018topological,nenno2020axion,mellnik2014spin,he2017tailoring,moghaddam2021prl,Hughes2022topologicalspintronics}. 
MTIs are characterized by a nontrivial $\mathbb{Z}_2$ invariant protected by the symmetry $S = \Theta \, T{1/2}$, which is the combination of time-reversal symmetry ($\Theta$) and a half-unit-cell translation symmetry ($T_{1/2}$). The surface states perpendicular to the magnetic ordering behave as 2D massive Dirac fermions due to the breaking of $\Theta$ symmetry \cite{Moore2010MTI, Vanderbilt2018axion, Wang2019axion, PhysRevLett.120.056801, liu2020robust, Sekine2021, Tan2023PRL}. 
While this phenomenon naturally occurs in ferromagnetic systems, it also arises in layered AFM materials where the magnetic moments align ferromagnetically within each layer in a direction perpendicular to the layers and alternate antiferromagnetically between adjacent layers \cite{bernevig2022progress,PhysRevX.9.041038}. 
As a result of the nonvanishing Berry curvature, these surfaces become 2D Chern insulators with a half-quantized Hall conductance. Thus, depending on the relative orientation of the magnetization on the top and bottom surfaces, MTIs can exhibit either a quantum anomalous Hall phase or an axion insulator phase \cite{bernevig2022progress, Tokura2019}.

\begin{figure}[t]
\centering
\includegraphics[width=.85\linewidth]{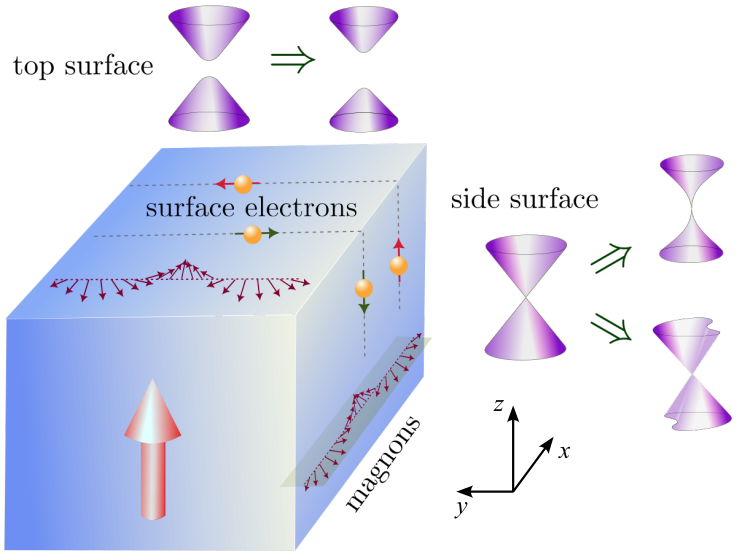}
\caption{\label{fig:sketch} 
Schematic illustration of magnon-mediated interactions effects on the topological surface states.
The surface states exhibit helical electronic modes, and magnons are represented in two orientations, perpendicular and parallel to the spin quantization axis (depicted by a prominent red arrow).
In the absence of interactions, the side surface is expected to remain gapless, while the top and bottom surfaces develop a gap if subjected to a finite exchange field $m_{\perp}$. Interactions lead to an increase in the gap amplitude at the top/bottom surfaces, whereas
on side surfaces we have significant velocity enhancement. Moreover, certain interaction limits can lead to strong spectral features similar to a massless gap opening.
}
\end{figure}

The pursuit of experimental realization of MTIs has met substantial success in candidate materials such as $\text{EuSn}_2\text{As}_2$ and $\text{MnBi}_{2n}\text{Te}_{3n+1}$  families \cite{otrokov2019prediction,PhysRevB.100.121104, PhysRevX.9.041038, PhysRevX.9.041039, PhysRevX.9.041040, Wueaax9989, Vidal2019, hu2020van, PhysRevB.101.161109, PhysRevB.101.161113, gordon2019strongly, PhysRevB.102.245136, Klimovskikh2020, xu2019persistent, PhysRevX.10.031013,Qihang2021route}.
Interestingly, the precise measurement of surface states dispersion and especially the magnetic gap 
have turned out to be challenging and still under debate from theoretical and experimental perspectives \cite{PhysRevLett.126.176403, hu2020realization, shikin2021sample,PhysRevB.103.L180403,facio2022}.
In a standard scenario, while symmetry breaking in magnetically ordered phases ensures the existence of a surface state Dirac mass gap, it is expected that the gap will vanish above the critical temperature. However, various experiments indicate that the gap persists even beyond the magnetic transition \cite{PhysRevX.9.041038, PhysRevX.9.041039, PhysRevX.9.041040}. Conversely, there are other findings in FM compounds or heterostructures, such as $\mathrm{MnBi_8Te_{13}}$, which exhibit substantial consistency with the aforementioned symmetry-breaking framework \cite{Chen2021, Rader2021, Edmonds2022, Macdonald2022}.

In this paper, motivated by the ongoing debate in the community and the overlooked interplay between magnetic excitations and electronic properties \cite{sudbo2021,Nogueira2021,lujan2022magnons}, we investigate the role of magnons in the surface states of MTIs. We demonstrate that magnons can mediate a long-range effective electron-electron interaction, leading to notable many-body effects on the spectral properties of the Dirac surface states, as illustrated in Fig. \ref{fig:sketch}. For the surfaces perpendicular to the magnetic ordering, there is a significant enhancement of the gap, which can exceed the non-interacting Dirac gap. For surfaces parallel to the magnetization, magnon-electron coupling causes either isotropic or anisotropic velocity renormalization, the latter accompanied by a tilting effect. In the limit of gapless magnons, isotropic velocity renormalization can asymptotically approach diverging Fermi velocity, similar behavior to \emph{massless gap opening} \cite{Cappelluti2008, Cappelluti2014,Rostami2018, Nagaosa2016weyl, Budich2019weyl}.
The resulting massless-gapped phase resembles properties of the Hatsugai-Kohmoto (HK) model \cite{HatsugaiKohmoto}, which has recently considered as an interesting exactly solvable playground to explore unconventional superconductivity \cite{Phillips2020,mai2024new}.

\section*{Results}
\noindent{\bf Surface electrons coupled to magnons}.
In the low-energy limit, the surface states are typically governed by the Dirac Hamiltonian
\begin{align}
    {\cal H}_{e} = \sum_{\bf k} 
    {\bm \psi}^\dag_{{\bf k}}
    \big(  \hbar v_F {\bf k} \cdot{\bm \sigma} + {\bm m}\cdot{\bm \sigma} \big)
    {\bm \psi}_{{\bf k}}, \label{eq:free-particle-H}
\end{align}
Here, ${\bm \psi}^\dag_{{\bf k}} =  \big(\psi^\dag_{{\bf k}\uparrow} ,\psi^\dag_{{\bf k}\downarrow}\big)$ represents the two-component creation field operator, and ${\bm \psi}_{{\bf k}}$ is the corresponding annihilation operator for electrons with momentum ${\bf k}$. The magnetic term ${\bm m}\cdot{\bm \sigma}$ at each surface with respect to its orientation decomposes into a Dirac gap contribution $m_\perp \sigma_\perp$ and a momentum shift term $m_\parallel \sigma_\parallel$, which can be gauged out for constant uniform magnetization. Therefore, we only need to consider the Dirac gap, caused by the perpendicular component of the magnetization to the considered surface. It is simply $m_z$ for the top surface in the $xy$ plane, while it vanishes for the side surfaces in the $xz$ or $yz$ plane, as depicted in Fig. \ref{fig:sketch}. The magnetic ordering of MTI can be effectively modeled with a Heisenberg Hamiltonian
\begin{align}
    {\cal H}_{\rm spin} = -\frac{1}{2} \sum_{\langle ij \rangle} J \: {\bf S}_{\rm loc}^i \cdot {\bf S}_{\rm loc}^{j} -K  \sum_{i} \big(S_{\rm loc}^{i,z}\big)^2.
\end{align}
This Hamiltonian includes Heisenberg interaction terms $J$ for spins at neighboring locations $i$ and $j$, and a magnetic anisotropy parameter $K$ \cite{HolsteinPrimakoff,Duine2010Magnon}. The spin operators in terms of Holstein-Primakoff (HP) transformation are given by
\begin{align}
    S_{\rm loc}^{i,+} \approx \sqrt{2s}\:b_i^\dag
    ~, \quad
    S_{\rm loc}^{i,-} \approx \sqrt{2s} \: b_i 
    ~, \quad
    S_{\rm loc}^{i,z}=s-b_i^\dag b_i.
\end{align}
Utilizing the HP transformation, and using Fourier transformation
$b_i = N^{-1/2}\sum_{\bf q} e^{i{\bf q}\cdot {\bf R}_i}  \: b_{\bf q} $,
we obtain the dispersion relation of magnons (spin waves) in the system as
\begin{align}
    {\cal H}_{\rm m} = \sum_{\bf q} \omega_{\bf q} b^\dag_{\bf q}b_{\bf q},
\end{align}
with the effective low-energy dispersion relations $\omega_{\bf q} = \Delta+D|{\bf q}|^2   a_0^2$ and $\omega_{\bf q} =  \sqrt{{\Delta}^2+D^2|{\bf q}|^2a_0^2} $ in the FM and AFM phases, respectively \cite{Kittel1963,prabhakar2009spinwave,Sudbo2019}. 
Assuming a square lattice, the gaps and dispersion coefficients for the FM and AFM cases are given by
\begin{align}
   & \Delta = \left\{\begin{array}{cc}
      2s\, K~,   & {\rm (FM)} \\[5pt]
       2s \sqrt{K(K+4J)} ~,   
      & {\rm (AFM)} 
    \end{array}\right. \\[5pt]
    & D= 2sJ,
\end{align}
where $s$ represents a spin of local moments as in HP transformation, and $a_0$ denotes the lattice spacing \cite{Sudbo2019,sudbo2020}. Note the different physical units of $D$ and $\Delta$ for the FM and AFM cases \footnote{The reason for this choice, which will be clear later, is to define a single form for the screening length.}.

There exists an $sd$ exchange coupling between itinerant electrons of the surface states and the local moments. This coupling can generally be expressed as
\begin{align}
    {\cal H}_{\rm sd} = \sum_{\bf q} {\bf s}_{\bf q}  \cdot {\mathbb J}_{\rm sd} \cdot {\bf S}_{\rm loc} (-{\bf q}),
\end{align}
where the itinerant electrons' spin density operator is given by ${\bf s}_{\bf q} = N^{-1/2}\sum_{\bf k}{\bm \psi}_{\bf k+q}^\dag  {\bm \sigma}{\bm \psi}_{\bf k}$ \cite{SM}. Here, ${\mathbb J}_{\rm sd}$ denotes the matrix of coupling strength, with components
\begin{align}
    \mathbb {J}_{\rm sd} = \begin{pmatrix}
        \mathbb {J}^{\rm xx}_{\rm sd} &  \mathbb {J}^{\rm xy}_{\rm sd}\\[5pt]  \mathbb {J}^{\rm yx}_{\rm sd} & \mathbb {J}^{\rm yy}_{\rm sd}    
    \end{pmatrix},
\end{align}
assuming a magnetization along the $z$ direction and neglecting the component of magnons parallel to the magnetization \footnote{This approximation is equivalent to ignoring the finite magnon population, justified at low temperatures, $\langle b_q^\dag b_q\rangle\approx 0$.}.
In general, the full $3\times3$ matrix ${\mathbb J}_{\rm sd}$ can be decomposed into an isotropic Heisenberg, a Dzyaloshinskii–Moriya (DM), and an anisotropic Ising term, as shown in Ref. \cite{Bellaiche2020}.
Substituting the result of the HP transformation for local spins in the $sd$ Hamiltonian, we obtain the electron-magnon interactions 
\begin{align}
    {\cal H}_{\rm em} = \sum_{\bf q} g_{\rm em} \: 
    \big(  {\bf s}_{\bf q} \cdot {\hat{\bf n}}\: b_{\bf q}^\dagger +
    {\bf s}_{\bf q} \cdot {\hat{\bf n}}^\ast \: b_{\bf -q} \big),
\end{align}
where the electron-magnon coupling strength is given by $g_{\rm em}=|{\mathbb J}_{\rm sd}|$, determined by the determinant of the $sd$ coupling matrix. In the coordinate system where magnetization is aligned along $z$, the complex unit vector ${\bf n}$ is expressed as
\begin{align}
   {\hat{\bf n}} 
   = \frac{1}{|{\mathbb J}_{\rm sd}|}
   \left[
   \big(
   {\mathbb J}_{\rm sd}^{\rm xx} + i \,{\mathbb J}_{\rm sd}^{\rm xy}
   \big) \hat{\bf x}+ 
   \big(
   {\mathbb J}_{\rm sd}^{\rm yx} + i \,{\mathbb J}_{\rm sd}^{\rm yy}
   \big) \hat{\bf y}
   \right].  
\end{align}
In special cases when the $sd$ exchange consists of either isotropic Heisenberg, anti-symmetric (DM), symmetric (Kitaev), or Ising types, the unit vector has a simple expression as summarized in the following:
\begin{equation}
\hat{\bf n} = \left\{
    \begin{array}{cc}
    \hat{\bf x} + i\,\hat{\bf y} &  \text{Heisenberg}\\
    \hat{\bf y} - i\,\hat{\bf x}  &  \text{anti-symmetric (DM)}\\
    \hat{\bf y} + i\,\hat{\bf x}  &  \text{symmetric (Kitaev-like)}\\
    \hat{\bf x} \:{\rm or}\: \hat{\bf y} & \text{Ising}
    \end{array}
    \right.\,.
\end{equation}
Combining all elements, the complete Hamiltonian of the surface states coupled to magnetic excitations is given by $  {\cal H} =  {\cal H}_e+ {\cal H}_m +  {\cal H}_{\rm em}$.

\begin{figure}[t]
\centering
\includegraphics[width=.9\linewidth]{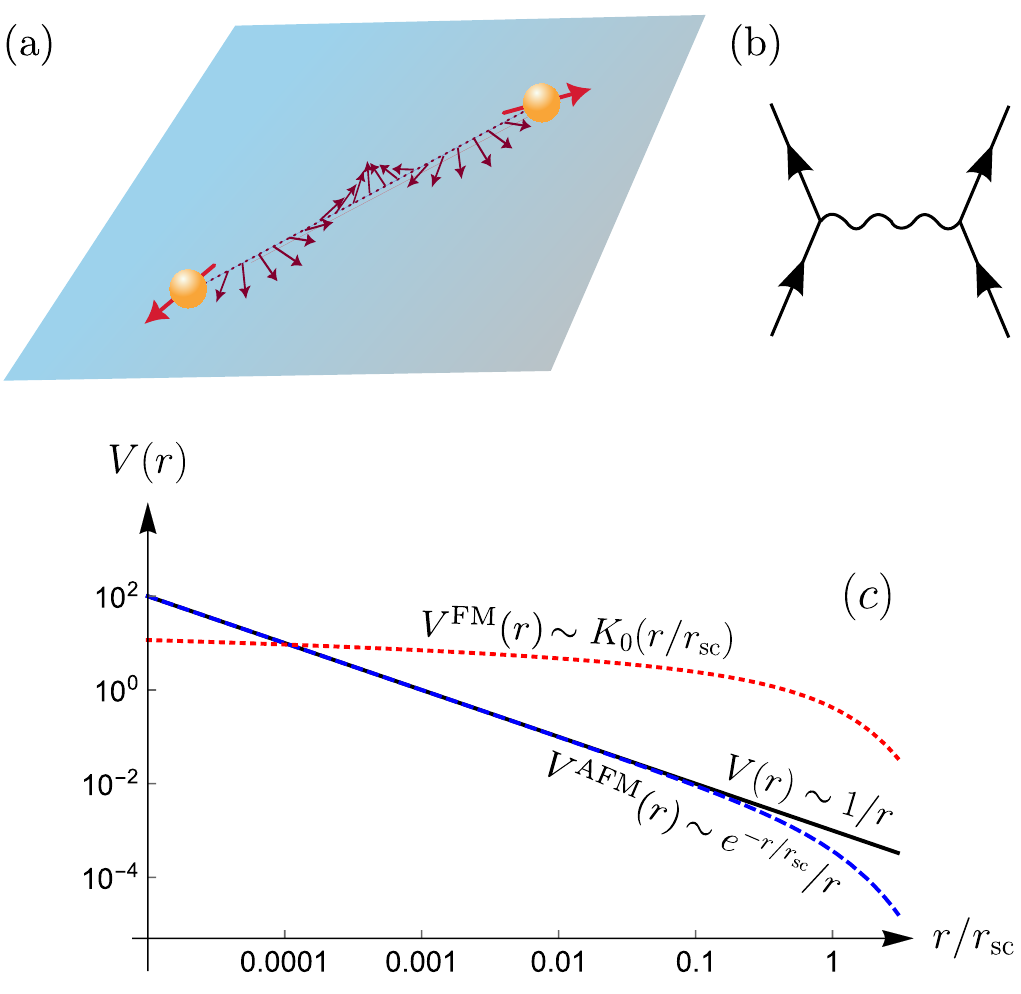}
\caption{\label{fig:interaction} 
Magnon-mediated interaction between itinerant electrons within surface states
(a) The schematic depiction illustrating how magnons
can mediate an effective 
(possibly long-range) interaction among surface-state electrons. (b) A corresponding Feynman diagram is presented, featuring straight arrows indicating the electron propagator and a wiggly line representing the magnon propagator. (c) The profile of the magnon-mediated interaction potential $V(r)$ , is depicted on a log-log scale against the interparticle distance 
$r$. The comparison includes FM (dotted red curve) and AFM (dashed blue curve) systems, juxtaposed with the standard Coulomb potential 
$V(r)\sim 1/r$. The effective screening length is expressed as
$r_{\rm sc}=a_0\sqrt{D/\Delta}$ and $r_{\rm sc}=a_0{D/\Delta}$ for FM and AFM, respectively. The numerical values along the $y$-axis are presented in arbitrary units.
}
\end{figure}

\vspace{2mm}
\noindent{\bf Magnon-mediated fermionic interactions}. By treating the electron-magnon couplings perturbatively, a well-justified approach in realistic magnetic materials, we can integrate out the magnon part of the Hamiltonian, resulting in effective interactions among electrons. This process, schematically illustrated in Fig. \ref{fig:interaction}, can be mathematically executed using the canonical transformation method, as shown in Supplementary Materials (SM) \cite{SM}. In the low-temperature limit, where magnetic excitations are nearly frozen, the effective interaction between surface states electrons is obtained as
\begin{align} 
    {\cal H}_{\rm int} &\approx \frac{-1}{2} \sum_{\bf q}\frac{g_{\rm em}^2}{\omega_{\bf q}}
    \sum_{\mu\nu} n_\mu n_\nu^\ast 
    s^\mu_{\bf q} s^\nu_{\bf -q}
    \nonumber\\
    &
    =
    \frac{-1}{2N} \sum_{\bf q,k,k'}\frac{g_{\rm em}^2}{\omega_{\bf q}}
    {\bm \psi}_{\bf k+q}^\dag  {\bm \sigma}\cdot\hat{\bf n}{\bm \psi}_{\bf k}\:
    {\bm \psi}_{\bf k'-q'}^\dag  {\bm \sigma}\cdot\hat{\bf n}^\ast  {\bm \psi}_{\bf k'}
    .\label{eq:eff-interaction}
\end{align}
The matrix structure of this interaction is rich, offering a versatile situation where different forms of interaction can be acquired by manipulating the details of the underlying electron-magnon coupling, which may vary with the types of magnetic materials and their orderings.

In general, electron-magnon coupling can have some momentum dependence as $g_{\bf q}$, which is disregarded here when starting from a short-range $sd$ coupling as the underlying mechanism of electron-magnon interaction. Now, due to the magnon dispersion, the resulting fermionic interaction matrix element in Eq. \eqref{eq:eff-interaction} becomes $V^{\rm FM}_{q}= g_{\rm em}^2/(\Delta+D (a_0q)^2)$ and
$V^{\rm AFM}_{q}= g_{\rm em}^2/({\Delta}^2+D^2 (a_0q)^2)^{1/2}$ for FM and AFM cases, respectively.
The more interesting result is for FM system which despite its resemblance to a screened Coulomb interaction in three spatial dimensions, it significantly differs from a Coulomb-like interaction as it acts between essentially 2D electrons. The real-space form of this interaction is given by the modified Bessel function of the second kind $V^{\rm FM}(r) = V_0 K_0(r/r_{\rm sc})$, with $V_0=g^2_{\rm em}/(2\pi D)$ and an effective screening length $r_{\rm sc} = a_0 \sqrt{D/\Delta}$ determined by the magnon gap. Beyond this length, the potential $V^{\rm FM}(r)$ exponentially decays. Unlike the ordinary Coulomb interaction, where the screening parameter is determined by the electronic density of states (DOS) at the Fermi level, here the effective screening is controlled by the magnonic gap $\Delta$ and is completely independent of DOS. In the limit of a vanishingly small magnon gap, the screening length diverges, and the interaction becomes long-range with very weak logarithmic dependence. The resulting long-range effective interaction between electrons when the magnons are gapless, asymptotically approaches an infinite-range interaction, resembling the exactly-solvable model of Hatsugai and Kohmoto \cite{HatsugaiKohmoto}.
In AFM case though, as we have shown in SM \cite{SM},
the real space form of the interaction turns out to be identical to fully screened Yukawa type form $V^{\rm AFM}(r)\propto e^{-r/r_{\rm sc}}/r$ for which we do not expect a radically interesting renormalization effects on the dispersion of topological surface states. This is in part understandable 
from the fact that due to the difference in their low-energy dispersion relations, FM magnons density of states are typically much higher than AFM case as it can be seen from their explicit expressions 
\begin{equation}
    \rho(\omega)= 
    \left\{\begin{array}{cc}
    \frac{\sqrt{\omega-\Delta}}{D^{3/2}}~,   & {\rm (FM)} \\
    & \\
    \frac{\omega \sqrt{\omega^2-\Delta^2}}{D^3} ~.   & {\rm (AFM)} 
    \end{array}\right.    
\end{equation}
In particular when the magnetic anisotropy is negligible and thus the magnon gap $\Delta$, we have $\rho_{\rm AFM}/\rho_{\rm FM} = \omega/D$
which becomes very small for low-energy magnons $\omega/D\ll 1$.

Here, our main focus is on the limit of unscreened or weakly screened effective interaction as it is possible in FM case, which will be discussed next. But before that, let us highlight the basic features of the opposite limit, characterized by a large magnon gap leading to an almost flat dispersion $\omega_q \sim {\rm cte}$ for magnons. In this case, the effective interaction is strongly screened, acquiring a momentum-independent coupling constant and giving rise to a short-range interaction. The resulting full Hamiltonian in this limit resembles the Thirring model, a simplistic quantum field theory for interacting fermions \cite{THIRRING195891,Cirac-Thirringmodel}. Moreover, it constitutes a special case of the \emph{Gross-Neveu model}, first introduced as a toy model in quantum chromodynamics \cite{GrossNeveu,Zinn-Justin1991,Zinn-Justin2003}. This model has recently regained interest, particularly within the condensed matter community \cite{schnetz2004phase,Herbut2006,Herbut2014grapheneRG,Sorella2016,Scherer2017,Lauchli2019Dirac,Adam2018,Vishwanath2018,Swingle2019Gross-Neveu-Yukawa,bentov2015fermion,Gracey2018}. The connection to the Gross-Neveu model \footnote{More precisely, the so-called chiral Heisenberg Gross-Neveu model with the Lagrangian ${\cal L}=i\bar{\psi}_i\bar{\partial}\psi_i +g^2
(\bar{\psi}_i\bm{\sigma}\psi)_i\cdot (\bar{\psi}_i\bm{\sigma}\psi_i)
$ (see, e.g., \cite{Gracey2018}).} becomes more profound when dealing with situations with more than one Dirac cone, a typical scenario in Weyl/Dirac materials. Indeed, the role of magnon-induced interactions in magnetic Weyl and Dirac semimetals is an interesting problem on its own, which will be studied elsewhere.

 \vspace{2mm}
\noindent{\bf Interaction effect on surface states}. We are now prepared to investigate the influence of the interactions given by Eq. \eqref{eq:eff-interaction} on the dispersion of topological surface states. In the weak coupling limit, the primary effect of interactions is represented by the self-energy at the one-loop approximation, given by
\begin{align}
    \bm{\Sigma}({\bf k})  = \bm{\sigma}\cdot \hat{\bf n}\, \bm{\Xi}({\bf k})\,{\bm\sigma}\cdot \hat{\bf n}^\ast .  
\end{align}
Here, $\bm{\Xi}({\bf k})$ is defined as
\begin{align} \label{eq:selfenergy}
    \bm{\Xi}({\bf k}) = -\frac{1}{N}\sum_{\bf k'} \frac{g_{\rm em}^2}{\omega_{\bf k-k'}} \,i\int \frac{d\omega}{2\pi} 
     {\bf G}({\bf k}',i\omega) ,
\end{align}
which corresponds to the exchange contribution \cite{mahan2013many,Rostami2018}. 
The bare Green's function ${\bf G}({\bf k},i\omega)$ of the surface states is obtained from 
the Hamiltonian \eqref{eq:free-particle-H} and is given by
\begin{align}\label{eq:bare-green}
    {\bf G}({\bf k},i\omega) = - \frac{i\omega +  \hbar v_F{\bf k} \cdot {\bm\sigma}  + 
    m_\perp \sigma_\perp }{\omega^2 +   \hbar^2 v_F^2|{\bf k}|^2 +m_\perp^2}.
\end{align}
As detailed in the SM \cite{SM}, we can precisely derive the self-energy expression
\begin{align}\label{eq:selfenergy-short}
    \bm{\Xi}({\bf k})  =   
    -f_3(k) \:{\hbar}v_F{\bf k} \cdot {\bm\sigma} - f_1(k) \: m_\perp \sigma_\perp.
\end{align}    
Here, the momentum-dependent renormalization functions are given by the expression (the graphical illustration of $f_n$ functions are given in \cite{SM})
\begin{align}
 f_n(\tilde{k})= \int_0^1 \frac{dx}{4\pi} \frac{ x^{n/2}(1-x)^{-1/2}}{ \sqrt{x \alpha^2 + x(1-x)\tilde{k}^2 +(1-x)\tilde{m}_{\perp}^2}},
\end{align}
which depends on the momentum, as well as
the magnon and electron gaps. Please note that 
 $\tilde{k}=k/k_\circ$ and  $\tilde{m}_{\perp}= m_{\perp}/(\hbar v_F k_\circ)$ are scaled with 
$k_\circ$ denoting a momentum scale that we set it equal to $1/a_0$. Consequently, the parameter $\alpha = 1/(k_\circ r_{\rm sc})  = \sqrt{\Delta/D},\text{~and~} {\Delta/D}$ (for FM and AFM, respectively) denotes the dimensionless scale of the screening length $r_{\rm sc}$ or equivalently, the magnon gap $\Delta$.

\section*{Discussion}

Since the self-energies are purely real, there is no level broadening, and the corresponding many-body effects 
only yield a renormalization of the energy dispersion, given by 
$\varepsilon_{\bf k,\pm}={\rm diag}\big[{\cal H}_e({\bf k})+\bm{\Sigma}({\bf k})\big]$. 
The self-energy, as given by Eq. \eqref{eq:selfenergy-short}, results in three distinct manifestations 
of band renormalization effects whose strengths determined by the functions $f_{1,3}(\tilde{k})$. First, there is gap renormalization, characterized by the term proportional to $m_\perp \sigma_z$ for top or bottom surfaces. 
The second effect is momentum-dependent velocity renormalization, which itself divides into isotropic 
and anisotropic velocity renormalization, as discussed below. 
Interestingly, due to the spin-dependent matrix structure of the effective interactions \eqref{eq:eff-interaction}, 
and considering that only the top surface has a Dirac gap $m_\perp$ in the non-interacting limit, the gap and 
velocity renormalization effects are only present for the top and side surfaces, respectively.
Finally, we can also have a momentum-dependent tilting of Dirac cone as we will see later on.

\begin{figure}[t]
\centering
\includegraphics[width=.99\linewidth]{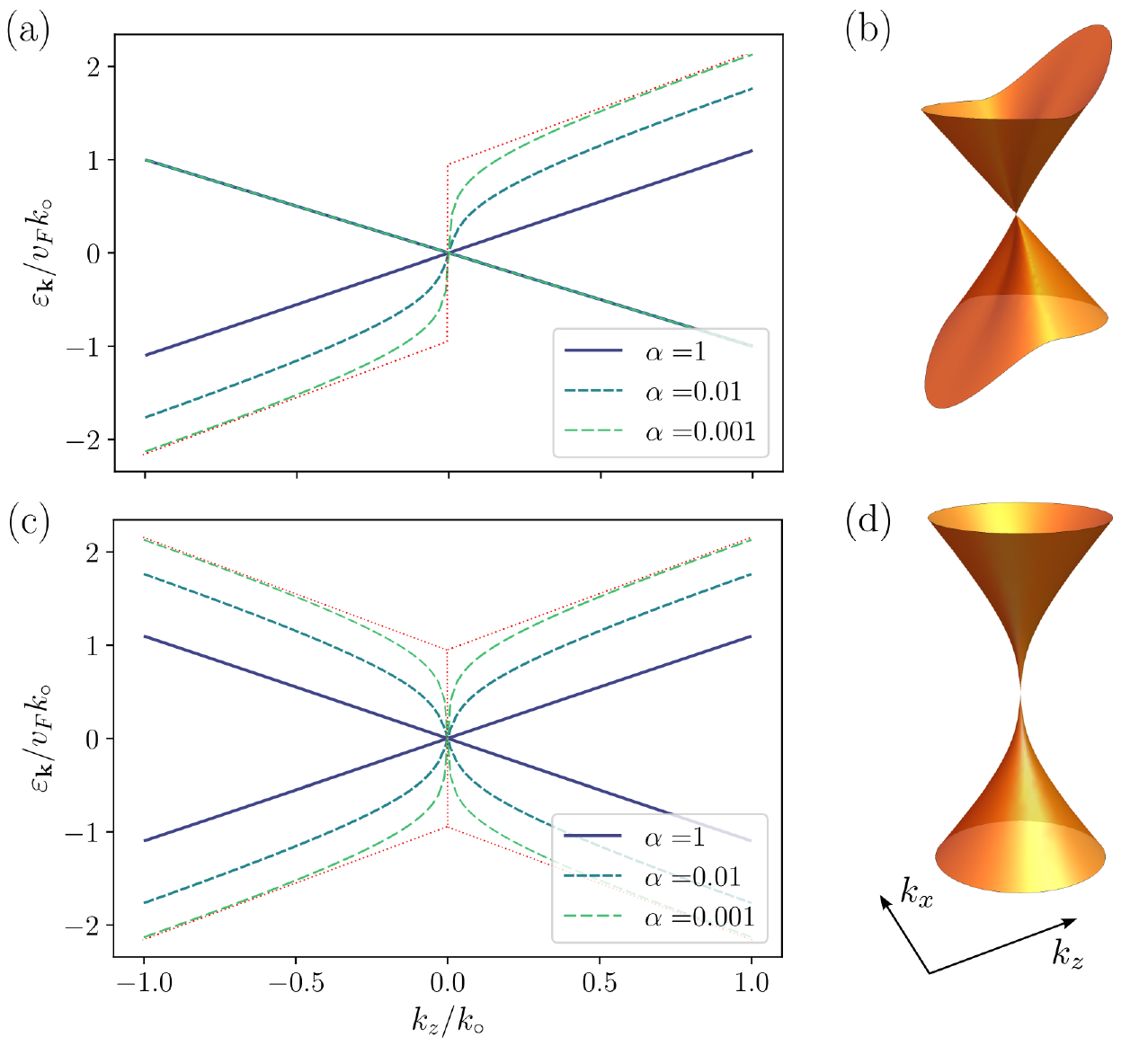}
\caption{\label{fig:spectrum} 
Interaction-induced changes in the dispersion relation at the vicinity of Dirac point for side surfaces.  
(a) and (b) show the spectrum for anisotropic case caused by the self-energy corrections in Eq. \eqref{eq:Sigma_aniso}, while (c) and (d) correspond to the isotropic case governed by the self-energy in Eq. \eqref{eq:Sigma_iso}.
In (b) and (d) we have set $\alpha=0.01$ with $\alpha$ 
being related to the magnon gap.
The velocity renormalization in the vicinity of Dirac point further increases by decreasing $\alpha$ and asymptotically diverges for $\alpha\to 0$.
Red dotted lines in (a) and (c) are shown as eye-guide
to illustrate the connection to massless-like gap opening.
}
\end{figure}

Let us first consider the side surfaces where, due to the absence of a Dirac gap in the noninteracting limit, 
only the first term of Eq. \eqref{eq:selfenergy-short} is relevant to them. In the cases of isotropic Heisenberg-, DM-, and Kitaev-type electron-magnon couplings, the self-energy corrections are expressed as 
\begin{align}
  {\bm\Sigma}_{\rm aniso}(k_x,k_z)|_{\rm side} = { 
  \frac{g^2_{\rm em}}{2D}f_3(\tilde{k})\: \tilde{k}_z} \big(\pm\sigma_0 +\sigma_z\big),  \label{eq:Sigma_aniso}
\end{align}
where the $'-'$ sign applies for the Kitaev-type coupling and the $'+'$ for the two others. This implies that solely the $z$-component of velocity within one of the two dispersion branches undergoes renormalization, 
leading to a combination of velocity change and {\em tilting} effects associated to $\sigma_z$ and $\sigma_0$ terms, respectively. 
These features are shown in Fig. \ref{fig:spectrum}(a), illustrating the effective band dispersion with respect to 
momentum $k_z$ when $k_x=0$. Consequently, the complete dispersion as a function of both momenta $(k_z,k_x)$ unveils a 
remarkably intriguing anisotropic spectrum, as demonstrated in Fig. \ref{fig:spectrum}(b).

Interestingly, with a distinctly anisotropic Ising-type electron-magnon coupling, denoted by $\hat{\bf n}_{\rm I}=\hat{\bf y}$ 
for the side surface in the $xz$ plane, an entirely isotropic self-energy given by
\begin{align}
  \bm{\Sigma}_{\rm iso}(k_x,k_z) |_{\rm side} = 
  \frac{g^2_{\rm em}}{2D}f_3(\tilde{k}) \: \tilde{\bf k}\cdot{\bm \sigma} \label{eq:Sigma_iso}
\end{align}
emerges, 
which gives rise to an isotropic velocity renormalization. The modified dispersion relations are illustrated 
in Fig. \ref{fig:spectrum}(c) and (d) for surface states at the side surfaces, assuming the Ising form for 
electron-magnon coupling. The results clearly indicate that the most significant changes in the dispersion occur in
the vicinity of momenta $k r_{\rm sc}\lesssim 1$, with the strength of renormalization dictated by
the dimensionless parameter $\alpha=1/(k_\circ r_{\rm sc})$. This is evident when considering the small momentum limit 
$k \to 0 $ in the presence of a finite screening length $r_{\rm sc}$, which results in a normalized velocity
\begin{align}
 \frac{v_*}{v_F} = 1+\frac{g^2_{em} }{6\pi D } \frac{r_{\rm sc}}{\hbar v_F}
\end{align}
by taking the small momentum limit of the renormalization function 
$f_3(\tilde{k}\to 0)\to 1/(3\pi\alpha)= k_\circ r_{\rm sc}/(3\pi)$. 
The impact of the many-body interaction is weak or moderate when $\alpha \lesssim 1$, corresponding to a short-range interaction potential with relatively small $r_{\rm sc}$. 

However, the normalized velocity grows linearly with $r_{\rm sc}$ and, notably, diverges in the limit of an infinitely large screening length ($r_{\rm sc}\to \infty$ or equivalently $\alpha \to 0$). This limit can be also understood by considering a finite nonzero momentum and first taking the limit $\alpha \to 0$ 
of the renormalization function $f_3 $, which is found to be
$\lim_{\alpha\to 0}\,f_3 = I k_\circ/k$ with $I=\int^1_0 \frac{dx}{4\pi} x(1-x)^{-1}$ being a singular integral. Assuming a regularised value of integral $I$ implies that $f_3   {\bf k}\cdot {\bm \sigma}\to I k_\circ \hat {\bf k}\cdot {\bm \sigma}$ which resembles massless gap term \cite{Rostami2018}. 
This scenario aligns with the previously mentioned HK physics, 
resulting in the appearance of a rigid shift of Dirac cones in opposite directions in energy as depicted by dotted red lines in Fig. \ref{fig:spectrum}(a) and (c).

Now, we turn to the surfaces in the $xy$ plane perpendicular to the magnetization, for which the self-energy becomes
\begin{equation}
{\bm\Sigma}(k_x,k_y)|_{\rm top} = \frac{g^2_{em}}{2D}   \frac{m_{z}}{\hbar v_F k_\circ}\, f_1(\tilde{k}) \big(\pm\sigma_0 +\sigma_z\big), \label{eq:Sigma_top}
\end{equation}
for the three forms of electron-magnon coupling matrix. 
Note that $m_\perp$ has been we have substituted with $m_{z}$ for the surfaces parallel to the $xy$ plane.  
In Eq. \eqref{eq:Sigma_top}, the $'+'$ sign corresponds to the isotropic Heisenberg and DM cases, while the $'-'$ sign corresponds to the Kitaev-type form of electron-magnon coupling. Remember that the polarization vectors for the above three cases are given by
$\hat{\bf n}_{H}=\hat{\bf x}+i\hat{\bf y}$ and $\hat{\bf n}_{\rm DM}=\hat{\bf y}-i\hat{\bf x}$, 
respectively.
By plugging Eq. \eqref{eq:selfenergy-short} into 
Eq. \eqref{eq:selfenergy} for the two polarization vectors, we can easily obtain the self-energy \eqref{eq:Sigma_top}.
Subsequently, for surface states in the $xy$ planes, 
we find that the main effects are a renormalized Dirac gap 
${m}_* = m_z \left[1+\frac{g^2_{em}}{2D(\hbar v_F k_\circ) } f_1(\tilde{k})\right]$
and a constant shift in energy. 
In the vicinity of the Dirac point ${\bf k}=0$, we find that  
$f_1({\bf k}=0)\to 1/(2\pi\alpha)$
and $f_1({\bf k}=0)\to 1/(2\pi \tilde{m}_z) \ln(\tilde{m}_z/\alpha)$ for the regimes of relatively small noninteracting gap ($\tilde{m}_z\ll \alpha$) and large gap ($\tilde{m}_z\gg \alpha$), respectively.
Hence, the renormalized gap for these two cases read
\begin{align}
\frac{{m}_*}{m_z} \approx \left\{
\begin{array}{cc}
    1+ \frac{g^2_{em}}{4\pi D } \frac{r_{\rm sc}}{\hbar v_F}  ,   & 
    r_{\rm sc} \,  \ll  \frac{\hbar v_F}{m_z}
    \\[5pt]
    1+\frac{g^2_{em}}{4\pi D m_z }
\,\ln(  m_z \, \frac{r_{\rm sc}}{\hbar v_F}
)  ,  &   r_{\rm sc} \,  \gg  \frac{\hbar v_F}{m_z}
\end{array} \right.
\end{align} 
This result indicates that as the magnon gap decreases, which also reduces $\alpha$, the renormalization becomes more significant in both of the limits above. On one hand, when the bare electronic gap and the magnon gap simultaneously approaches zero (i.e. $\alpha \sim \tilde m_z\to 0$), we can expect a regime where $m_\ast \propto m_z/2\pi\alpha)$ remains finite. This implies an interaction-induced gap-opening that requires a spontaneous breaking of the time-reversal symmetry.
On the other hand, when the bare Dirac gap is large ($m_z\gg v_F k_{\circ} \alpha$), the renormalization effects show a logarithmic dependence $\ln (m_z/\hbar v_F k_{\circ} \alpha)\equiv \ln (m_z r_{\rm sc}/ \hbar v_F)$ which diverges in the limit of $r_{\rm sc}\to \infty$ corresponding to gapless magnons. This logarithmic divergence arises from the strong infrared feature of the effective interaction coupling matrix element $g_{\rm em}^2/\omega_{\bf q}$ in the absence of a magnon gap. Therefore, it suggests that the massive Dirac liquid is prone to instabilities in the presence of these strong interactions, which warrants separate investigation in future studies.

\vspace{2mm}
\noindent{\bf Relevance to real materials}.
We have seen that magnons can mediate effective interactions among electrons 
with quite different behavior in FM and AFM TIs.
While in FM case we have a strong Coulomb-like 
interaction which approaches an infinite-range for gapless magnons at the level of one-loop self-energy,
for AFM we always get a highly screened Yukawa type  interactions. This readily can suggest that different forms of magnon dispersion and subsequently the effective magnon-mediated interactions for the
two different magnetic ordering can potentially be 
another source for the differences in their surface states. 

One of the most studied MTIs is stoichiometric $\mathrm{MnBi_2Te_4}$ with an $A$-type AFM ordering, where local moments are ferromagnetically aligned in the perpendicular direction within the Mn planes but antiferromagnetically ordered between adjacent layers. For this type of MTIs, as an approximation, we can assume that magnons contributing to the top/bottom surfaces are more like FM magnons, while for the side surfaces, AFM magnons have the major contribution. This means that in $\mathrm{MnBi_2Te_4}$, only the surfaces perpendicular to the magnetic ordering are significantly influenced by the magnon-mediated interaction effects studied here, since effective interactions due to the AFM magnons are much weaker. In contrast, our results for the side surface states are mostly relevant in fully ferromagnetically-ordered MTIs, such as $\mathrm{MnBi_8Te_{13}}$. In particular, MTIs with FM order, in which magnetic anisotropy is weak or absent and thus have a small magnon gap, are the best candidates to experimentally explore our predictions for many-body effects associated with magnons on the topological surface states.

As mentioned earlier, due to the long-range nature of the magnon-mediate interactions for gapless magnons,
the leading order self-energy calculation here coincides with the exact result for the case of infinite-range (HK) interaction \cite{HatsugaiKohmoto,Nogueira1996,Nagaosa2016weyl,Rostami2018}.
This consistency at the one-loop self-energy level, underscores the depth of the observed renormalization effects. Notably, the one-loop approximation utilized in this analysis approaches the accuracy of the infinite-range scenario, effectively governing the comprehensive many-body aspects, as demonstrated in prior investigations \cite{Nagaosa2016weyl, Rostami2018, Budich2019weyl}. Consequently, the profound renormalization of the dispersion relation induced by long-range magnon-mediated interactions should, at least in essence, persist in a more involved many-body analysis beyond one-loop approximation. 
To illustrate this point, by considering the extremely long-range limit where $V_{\bf q} \to \delta({\bf q})$, we show that the higher-order corrections to the self-energy vanish identically (see Section IV of SM \cite{SM}). This demonstrates that, while higher-order corrections remain finite for long-range interactions, the one-loop result remains as the leading term, particularly in the long-wavelength limit.
Nevertheless, a more careful study employing a renormalization group formulation can further substantiate this physical insight, akin to analogous studies conducted on Coulomb interactions in graphene \cite{Nagaosa2016, Adam2018}.

It should be mentioned that direct Coulomb electron-electron interactions and the associated screening or anti-screening due to many-body effects have been neglected in our analysis. However, our main findings remain qualitatively unchanged even when considering these Coulomb interaction effects. This persistence is due to the singularity of the magnon-mediated interaction at the long-wavelength limit, which cannot be eliminated by Coulomb screening effects. Additionally, we emphasize that our focus in this study is on the zero-temperature limit. This choice is justified by the low critical temperatures of most MTI materials, rendering finite-temperature effects negligible.

\section{Conclusions} 
In summary, we have found that magnon-mediated interactions play a crucial role in determining many-body effects on the dispersion relation of surface states in magnetic topological insulators. Specifically, we demonstrate that ferromagnetic magnons can induce strong fermionic interactions, making magnetic topological insulators promising systems for exploring intriguing yet elusive many-body effects. Notably, we observe gap enhancements and Fermi velocity divergences resembling spectra with massless-like gap openings on the top and side surfaces of ferromagnetic topological insulators (see Fig. \ref{fig:sketch}). Interaction-induced enhancement of Dirac gaps on surfaces perpendicular to the magnetic ordering, strengthen the topological protection of quantum anomalous Hall and axion insulator phases associated to the surface states.

Our findings regarding the coupling between surface electrons and bulk magnons offer insights into the electronic properties of topological surface states and may shed light on current discrepancies in photoemission spectroscopy of surface states. Furthermore, by exploring various methods for modifying the electronic dispersion of surface states, our findings can be applied to engineer spintronic applications of magnetic topological insulators, leveraging the interaction between magnons and electrons.

\vspace{0.5cm}
\noindent{\bf
Data availability statement.
} All the data that support the findings of this study are openly available in the \emph{Supplemental Materials} part.

\vspace{0.5cm}
\noindent{\bf Acknowledgement}. The authors acknowledge Emmanuele Cappelluti for helpful discussions.
HR acknowledges the
support of Swedish Research Council (VR Starting Grant
No. 2018-04252).
AGM acknowledges financial support from Jane and Aatos Erkko Foundation.

\bibliography{refs}
\end{document}


\title{Supplemental Materials: \\ Fate of surface gaps in magnetic topological insulators}

\author{Habib Rostami}
\address{Department of Physics, University of Bath, Claverton Down, Bath BA2 7AY, United Kingdom}

\author{Ali G. Moghaddam}
\address{Computational Physics Laboratory, Physics Unit, Faculty of Engineering and Natural Sciences, Tampere University, P.O. Box 692, FI-33014 Tampere, Finland}
\address{Helsinki Institute of Physics, FI-00014 University of Helsinki, Finland}

\date{\today}
\maketitle

In the supplementary materials, we provide detailed calculations that are not presented in the main text. The supplementary material is organized into five sections. In the first section, we present the conventions used in our derivations, especially for Fourier transformations. The second section is devoted to the derivation of magnon-mediated interactions using the canonical transformation technique. Subsequently, we obtain the explicit form of interactions in real space induced by low-energy magnons in both ferromagnetic and antiferromagnetic settings. In the antiferromagnetic case, we observe a strongly screened form of interaction, while in a ferromagnetic system, interactions can be unscreened when the magnon gap becomes very small. In the fourth section, we present comprehensive calculations for the electronic self-energy due to magnon-mediated interactions in the ferromagnetic case, where many-body effects are more pronounced. Finally, in the last section, we show the higher-order self-energy corrections vanishes in 
the infinite-range limit of interactions.
\section{Basic conventions}
\label{app:fourier}
We define the Fourier transforms for the local spin operators as:
\begin{align}
    \mathbf{S}(\mathbf{k}) &= \frac{1}{\sqrt{N}} \sum_{{\bf R}} e^{i \mathbf{k} \cdot {\bf R}} \mathbf{S}({\bf R}),
    \\
    \mathbf{S}({\bf R}) &= \frac{1}{\sqrt{N}} \sum_{\mathbf{k}} e^{-i \mathbf{k} \cdot {\bf R}} \mathbf{S}(\mathbf{k}),
\end{align}
in consistency with the completeness relations
\begin{align}
\delta_{{\bf k},{\bf k}'}&=\frac{1}{N} 
\sum_{{\bf R}} 
e^{i (\mathbf{k}-\mathbf{k}') \cdot {\bf R}} \\
\delta_{{\bf R},{\bf R}'}&=\frac{1}{N} 
\sum_{\mathbf{k}} 
e^{i \mathbf{k} \cdot ({\bf R}-{\bf R}')} \label{eq:completeness-2}
\end{align}

In the continuum limit for the momentum $k$
we can re-write Eq. \eqref{eq:completeness-2} as
\begin{align}
\delta_{{\bf R},{\bf R}'}&=\frac{1}{N}
\int \frac{d{\mathbf{k}}}{(2\pi/L)^2} 
e^{i \mathbf{k} \cdot ({\bf R}-{\bf R}')} 
\\
&=
a_0^2
\int \frac{d{\mathbf{k}} }{(2\pi)^2}
e^{i \mathbf{k} \cdot ({\bf R}-{\bf R}')} 
\end{align}
where we have used $a_0^2 = {L^2/N}$ that represents the unit cell area in two dimensions. So equivalently we can use the following convention for switching from discrete sum to integral over momentum: 
\begin{align}
    \frac{1}{N}\sum_{\bf k} \to a_0^2 \int   \frac{d{\mathbf{k}} }{(2\pi)^2}.
\end{align}
The above Fourier transformations for the itinerant electrons spin operators, can be expressed in terms of electron creation and annihilation operators as
\begin{align}
    \mathbf{s}(\mathbf{q}) &= \frac{1}{\sqrt{N}} \sum_{{\bf R}} e^{i \mathbf{q} \cdot {\bf R}} \mathbf{s}({\bf R}) = \frac{1}{\sqrt{N}} \sum_{{\bf R}} e^{i \mathbf{q} \cdot {\bf R}} {\bm\psi}^\dag({\bf R})\: {\bm \sigma}\: {\bm\psi}({\bf R})
    \nonumber\\
    &=
     \frac{1}{\sqrt{N}} \sum_{{\bf R}} e^{i \mathbf{q} \cdot {\bf R}} 
     \frac{1}{\sqrt{N}} \sum_{{\bf k}'}
     {\bm\psi}_{{\bf k}'}^\dag e^{-i {\bf k}' \cdot {\bf R}}
     \: {\bm \sigma}\:
     \frac{1}{\sqrt{N}} \sum_{{\bf k}}
     {\bm\psi}_{{\bf k}} e^{i {\bf k} \cdot {\bf R}}\nonumber\\
      &=
     \frac{1}{\sqrt{N}} 
      \sum_{{\bf k},{\bf k}'} \delta_{{\bf q}+{\bf k},{\bf k}'} \:
     {\bm\psi}_{{\bf k}'}^\dag  
     \: {\bm \sigma}\:
     {\bm\psi}_{{\bf k}} \nonumber\\
      &=
     \frac{1}{\sqrt{N}} 
      \sum_{{\bf k}}
     {\bm\psi}_{{\bf k}+{\bf q}}^\dag  
     \: {\bm \sigma}\:
     {\bm\psi}_{{\bf k}} 
\end{align}

The local interactions between itinerant electrons spin with localized spin (known as $sd$ exchange coupling) can be also written in $k$-space applying the Fourier transformations as the following
\begin{align}
    {\cal H}_{\rm sd} 
    &= \sum_{\bf R} {\bf s}({\bf R}) \cdot {\mathbb J}_{\rm sd} \cdot {\bf S}_{\rm loc} ({\bf R})\nonumber\\
    &= \sum_{\bf R} \frac{1}{N}\sum_{{\bf q},{\bf q}'}
    e^{-i{\bf q}\cdot {\bf R}} \:
    {\bf s}_{\bf q} \cdot {\mathbb J}_{\rm sd} \cdot {\bf S}_{\rm loc} ({\bf q}')\: e^{-i{\bf q}'\cdot {\bf R}} 
    = \sum_{{\bf q},{\bf q}'}
     \delta_{{\bf q},-{\bf q}'} \: 
    {\bf s}_{\bf q} \cdot {\mathbb J}_{\rm sd} \cdot {\bf S}_{\rm loc} ({\bf q}') 
    \nonumber\\
    &= \sum_{\bf q} {\bf s}_{\bf q}  \cdot {\mathbb J}_{\rm sd} \cdot {\bf S}_{\rm loc} (-{\bf q}).
\end{align}

\section{Derivation of magnon-mediated interactions}
\label{app:canonical}

We begin by considering a generic relation of canonical transformation in a perturbative was, which reads
\begin{align}
    \widetilde{\cal H} = e^{\cal O} {\cal H} e^{-\cal O}  = {\cal H} + [{\cal O},{\cal H}]+\frac{1}{2}[{\cal O},[{\cal O},{\cal H}]]+ \dots 
\end{align}
to find an effective electron-electron (e-e) interactions mediated by magnons.
To this end and in the spirit of Schrieffer-Wolff transformations, we require ${\cal H}_{\rm em}+[{\cal O} ,{\cal H}_m]=0$ which is fulfilled by the operator
\begin{align}
    {\cal O} 
    =  \sum_{q} \frac{g_{\bf q}}{\omega_{\bf q}} \: \big(  {\bf s}_{\bf q}\cdot {\hat{\bf n}}\: b_{\bf q}^\dagger - {\bf s}_{\bf q}\cdot {\hat{\bf n}}^\ast \: b_{\bf -q}  \big).
\end{align}
Then, considering the terms up to second-order in the electron-magnon coupling $g_{\bf q}$, we find {\color{black}{the transformed Hamiltonian}}   
\begin{align} \label{eq:H_eff_SW}
    {\color{black}{\widetilde{\cal H}}} =  {\color{black}{{\cal H}_e}+[{\cal O},{\cal H}_e]}+ \frac{1}{2}[{\cal O},{\cal H}_{\rm em}].
\end{align}
Since we want to obtain an effective interacting fermionic model,
another step is required to integrate out magnon degrees of freedom by 
taking the thermal average over the magnon fields (this is known as the Holstein approximation in the context of electron-phonon couplings and polaron physics).  
We first consider the first-order term
\begin{align}
   [{\cal O},{\cal H}_e] =  \sum_{{\bf k},{\bf q}} 
    \frac{g_{\bf q}}{\omega_{\bf q}}
    \psi_{\bf k}^\dag
    \big[{\bm \sigma}\cdot \hat{\bf n}, {\bf k}\cdot{\bm \sigma}\big] 
    \psi_{{\bf k}+{\bf q}}\: b_{\bf q}^\dag + h.c.,
\end{align}
which has a similar form to the original electron-magnon coupling but with a momentum-dependent coupling strength $g_{\bf q} \: \hbar v_Fk/\omega_{\bf q}$. 
Being linear in bosonic creation and annihilation operators, this
term vanishes by taking the thermal average since $\langle b_{\bf q}^\dag\rangle=\langle b_{\bf q}\rangle=0$.
The second term in the effective Hamiltonian Eq. \eqref{eq:H_eff_SW} is found as
\begin{align}
\frac{1}{2}[{\cal O},{\cal H}_{\rm em}]
=\frac{1}{2}
     \sum_{\bf q} \frac{g_{\bf q}^2}{\omega_{\bf q}}
   \Big(
  -\{
  {\bf s}_{\bf q}\cdot {\hat{\bf n}},
  {\bf s}_{\bf -q}\cdot {\hat{\bf n}}^\ast
  \}
  +
  2\:
   [
     {\bf s}_{\bf q}\cdot {\hat{\bf n}}
     ,
     {\bf s}_{\bf q}\cdot {\hat{\bf n}}^\ast]  \: b_{\bf q}^\dagger b_{-\bf q}
     \Big),
\end{align}
whose first term includes a fermionic interaction whereas the second term 
{\color{black}{after taking the thermal average over magnons}} yields a renormalization of the fermionic single-particle terms weighted by the magnons population $\langle b^\dag b \rangle$ \footnote{The fact that the second term represent a single-particle fermionic terms is followed from the commutation relation $[
     {\bf s}_{\bf q}\cdot {\hat{\bf n}}
     ,
     {\bf s}_{\bf q}\cdot {\hat{\bf n}}^\ast] = i \big(\hat{\bf n}\times \hat{\bf n}^\ast\big)\cdot {\bf s}_{\bf q}$}. 
In low-temperature limit where magnon excitations are almost frozen,
only the first term in above expression survives which results in {\color{black}{an effective
fermionic Hamiltonian
\begin{align} \label{eq:H_eff}
    {\cal H}_{\rm eff} =  {\cal H}_e + {\cal H}_{\rm int} ,
\end{align}
}}
with the interaction term  
\begin{align} 
{\cal H}_{\rm int} \approx -\frac{1}{2} \sum_{\bf q} \frac{g_{\bf q}^2}{\omega_{\bf q}}
  \{
  {\bf s}_{\bf q}\cdot {\hat{\bf n}},
  {\bf s}_{\bf -q}\cdot {\hat{\bf n}}^\ast
  \},\label{eq:eff-interaction-SM}
\end{align}
So that the effective interaction coupling is given by 
\begin{align}
    V_{\bf q} = \frac{g^2_{\bf q}}{\omega_{\bf q}}. 
\end{align}

\section{Effective interactions in real-space}
In this part we explore the form of magnon-mediated interactions in real space for both ferromagnetic and antiferromagnetic cases. For FM magnons using the low-energy dispersion relation $\omega_q=\Delta + Da^2_0q^2 = Da^2_0(r^{-2}_{\rm sc}+ q^2)$ with $r_{\rm sc}=a_0\sqrt{D/\Delta}$
and considering short-range electron magnon coupling 
with a constant momentum-independent coefficient $g_{q} = g_{\rm em}$, we have 
\begin{align}
    V({\bf r}) &= \frac{g_{\rm em}^2}{D}\int \frac{e^{-i{\bf q}\cdot {\bf r}}}{r^{-2}_{\rm sc} + q^2}  \frac{d^2 {\bf q} }{(2\pi)^2} 
= \frac{g_{\rm em}^2}{D\Gamma(1)} \int^\infty_0 d\tau  \int  e^{-i{\bf q}\cdot {\bf r}} e^{-\tau (r^{-2}_{\rm sc}+q^2)} \frac{d^2 {\bf q} }{(2\pi)^2}  
\nonumber\\
&= \frac{g_{\rm em}^2}{D} \int^\infty_0 d\tau e^{-\tau r^{-2}_{\rm sc}} \int  e^{-\tau  q^2-i{\bf q}\cdot {\bf r}}   \frac{d^2 {\bf q} }{(2\pi)^2}  
\nonumber\\
&= \frac{g_{\rm em}^2}{D} \int^\infty_0 d\tau e^{-\tau r^{-2}_{\rm sc}} 
\frac{\pi\tau^{-1}}{(2\pi)^2} e^{-\frac{r^2}{4\tau}}
\nonumber\\
&= \frac{g_{\rm em}^2}{4\pi D} 
\int^\infty_0 d\tau \tau^{-1} e^{-\tau r^{-2}_{\rm sc}} 
 e^{-\frac{r^2}{4\tau}}
\end{align}
Performing last integral over $\tau$ gives the Bessel function $K_0(x)$. Therefore 
\begin{align}
V({\bf r})= \frac{g_{\rm em}^2}{2\pi }  K_0(r/r_{\rm sc})
\end{align}
We can also obtain the asymptotic behaviors of this interaction in small and large distances;

\begin{align}
V({\bf r})= \left\{
\begin{array}{cc}
    \frac{g_{\rm em}^2}{2\pi D} \left\{ \ln (2)-\gamma - \ln (r/r_{\rm sc}) +\dots \right\}~ ,  &   r/r_{\rm sc}\ll 1 \\[5pt]
    \hspace{-2cm} \frac{g_{\rm em}^2}{2 D\sqrt{2 \pi }} \frac{e^{- r/r_{\rm sc}}}{ \sqrt{r/r_{\rm sc}}} + \dots ~ ,&  r/r_{\rm sc}\gg1
\end{array}\right.
\end{align}

\begin{figure}[t!]
\centering
\includegraphics[width=0.9\textwidth]{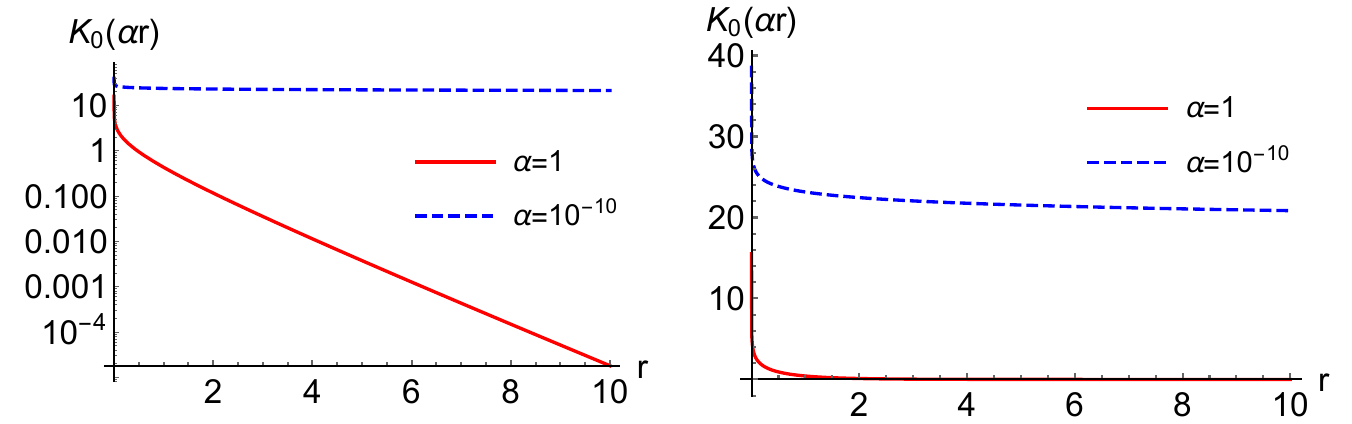}
\caption{left: log scale. right: linear scale. As seen for small $\alpha\to0$ we can approximate $V(r)$ as a constant potential in finite range of $r$.}
\end{figure}

The special case of $\alpha=0$ can be obtained explicitly as
\begin{align}
\int^\infty_0 d\tau \tau^{-1}   
 e^{-\frac{r^2}{4\tau}}=\int^\infty_0 dt t^{-1}   
 e^{-t\frac{r^2}{4}}
  = \frac{1}{\Gamma(0)} (\frac{4}{r^2})^0,
\end{align}
corresponding to infinite range interactions.

\subsection{Antiferromagnetic case}
For the antiferromagnetic we have $\omega_{\bf q} = \sqrt{\Delta^2+D^2 a^2_0 q^2} = D a_0\sqrt{r^{-2}_{\rm sc}+q^2}$ leading to 
\begin{align}
    V({\bf r}) &= \frac{g_{\rm em}^2 a_0}{D  }\int \frac{e^{-i{\bf q}\cdot {\bf r}}}{(r^{-2}_{\rm sc} + q^2)^{1/2}}  \frac{d^2 {\bf q} }{(2\pi)^2} 
= \frac{g_{\rm em}^2 a_0}{ D \Gamma(1/2)} \int^\infty_0 d\tau \tau^{-1/2} \int  e^{-i{\bf q}\cdot {\bf r}} e^{-\tau (r^{-2}_{\rm sc} +q^2)} \frac{d^2 {\bf q} }{(2\pi)^2}  
\nonumber\\
&= \frac{g_{\rm em}^2 a_0}{ D  \Gamma(1/2)} \int^\infty_0 d\tau \tau^{-1/2} e^{-\tau r^{-2}_{\rm sc}} \int  e^{-\tau  q^2-i{\bf q}\cdot {\bf r}}   \frac{d^2 {\bf q} }{(2\pi)^2}  
\nonumber\\
&= \frac{g_{\rm em}^2a_0}{D \Gamma(1/2)} \int^\infty_0 d\tau \tau^{-1/2} e^{-\tau r^{-2}_{\rm sc}} 
\frac{\pi\tau^{-1}}{(2\pi)^2} e^{-\frac{r^2}{4\tau}}
\nonumber\\
&= \frac{g_{\rm em}^2 a_0}{4\pi  D  \Gamma(1/2)} 
\int^\infty_0 d\tau \tau^{-3/2} e^{-\tau r^{-2}_{\rm sc}} 
 e^{-\frac{r^2}{4\tau}} 
 \nonumber\\
 &
 = \left(\frac{g_{\rm em}^2 a_0}{2\pi  D  r_{\rm sc}} \right) \frac{e^{-r/r_{\rm sc}}}{r/r_{\rm sc}} 
  = \left(\frac{g_{\rm em}^2 \Delta}{2\pi  D^2} \right) \frac{e^{-r/r_{\rm sc}}}{r/r_{\rm sc}} 
\end{align}

Therefore, for the case AF we obtain a screened inter-particle Coulomb interaction.

\section{Self-energy corrections}
In the short-range electron-magnon coupling limit where we have $g_{\bf q} = g_{\rm em} $ (subscript $s$ is used to indicates short range) independent of ${\bf q}$, 
the self-energy is given by
\begin{align} 
    {\bm\Sigma}({\bf k}) &=  {\bm\sigma}\cdot \hat{\bf n}
     \:{\bm\Xi}\:
     {\bm\sigma}\cdot \hat{\bf n}^\ast, \\
    {\bm\Xi}&=   - \frac{1}{N} \sum_{\bf k'} \frac{g_{\rm em}^2}{\Delta + D
    {\revise{a_0^2}}
    |{\bf k-k'}|^2 } i\int \frac{d\omega}{2\pi} 
     \: {\bf G}_0({\bf k}',i\omega) .
\end{align}
First, we note that the integration over frequency $\omega$, using the Green's function, Eq. (12) in the main text yields
\begin{align}
   i \int \frac{d\omega}{2\pi} {\bf G}_0 ({\bf k}',i\omega)=
    \frac{1}{2}\frac{ \hbar v_F{\bf k}' \cdot {\bm\sigma}  + m_z \sigma_{z}}{ \sqrt{\hbar^2 \hbar^2 v_F^2|{\bf k}'|^2 +m^2_z}}.
\end{align}
Therefore,
{\revise{
\begin{align} 
    {\bm\Xi}&=   -  \frac{g_{\rm em}^2}{2 D }
    \int a_0^2d^2{\bf k'}
    \frac{1}{\Delta/D + a_0^2|{\bf k-k'}|^2}
    \frac{ \hbar v_F{\bf k}' \cdot {\bm\sigma}  + m_z \sigma_{z}}{ \sqrt{\hbar^2 \hbar^2 v_F^2|{\bf k}'|^2 +m^2_z}}\nonumber\\
    &=   -  \frac{g_{\rm em}^2}{2 D }
    \int d^2\tilde{\bf k}'
    \frac{1}{\alpha^2 +|\tilde{\bf k}-\tilde{\bf k}'|^2 }
    \frac{ \tilde{\bf k}' \cdot {\bm\sigma}  + \tilde{m}_z \sigma_{z}}{\sqrt{|\tilde{\bf k}'|^2 +\tilde{m}_z^2}},
\end{align}
where we have introduced a parameter $\alpha = \sqrt{D/\Delta}$, then resealed the momentum
as ${\bf k} \to \tilde{\bf k}={\bf k}/k_\circ$
and the Dirac gap as $m_{z} \to \tilde{m}_{z}=m_{z}/({\hbar v_F k_\circ})$ in which  $k_\circ=1/a_0$.
The self-energy ${\bm \Xi}$ can be expressed 
as
\begin{align} \label{self-Xi}
    {\bm\Xi} &= -\frac{g_{\rm em}^2 }{2 D}
    \left[
    I_1(\tilde{\bf k}, \alpha ,\tilde{m}_z) \: \tilde{m}_z \sigma_{z}
    +I_{2,i}(\tilde{\bf k}, \alpha ,\tilde{m}_z) \: \sigma_i 
    \right]
    \nonumber\\
    &= {\color{black}\frac{- g^2_{em}}{2D(\hbar v_F k_\circ)  }}  \left[ I_1 (\tilde{\bf k}, \alpha ,\tilde{m}_z) \:  m_z \sigma_z + I_{2,i} (\tilde{\bf k}, \alpha ,\tilde{m}_z) \:  \hbar v_F k_\circ \sigma_i \right],
\end{align}
in terms of the following \emph{dimensionless} integrals
}}
\begin{align} 
 I_1({\bf k}, \alpha ,m)=\int \frac{d^2{\bf k'}}{(2\pi)^2} \frac{1}{\alpha^2 + |{\bf k-k'}|^2 } \frac{1}{ \sqrt{ |{\bf k}'|^2 +m^2}},
 \\
 I_{2,i}({\bf k}, \alpha ,m)= \int \frac{d^2{\bf k'}}{(2\pi)^2} \frac{1}{\alpha^2 + |{\bf k-k'}|^2 } \frac{ k'_i}{ \sqrt{ |{\bf k}'|^2 +m^2}}.
\end{align}
Now, to evaluate above integrals, we use the Feynman parametrization for arbitrary functions $h_1$ and $h_2$
\begin{equation} 
\frac{1}{h^\alpha_1 h^\beta_2}
=
\frac{\Gamma(\alpha+\beta)}{\Gamma(\alpha)\Gamma(\beta)}
\int^1_0 dx  
\frac{x^{\alpha-1} (1-x)^{\beta-1}}{\left [x h_1+(1-x) h_2 \right]^{\alpha+\beta}},
\end{equation}
Note the Euler's Gamma-function $\Gamma(z)=(z-1)!$. 
Let's first check the case of $I_1$ using the Feynman's integration trick:   
\begin{align} 
 \frac{1}{\alpha^2 + |{\bf k-k'}|^2 }  \frac{1}{( |{\bf k}'|^2 +m^2)^{1/2}}
 &= \frac{\Gamma(3/2)}{\Gamma(1)\Gamma(1/2)}\int_0^1 dx \frac{x^{1/2}(1-x)^{-1/2}}{\{x [\alpha^2 + |{\bf k-k'}|^2]+ (1-x) [|{\bf k}'|^2 +m^2]\}^{3/2}}
 \nonumber\\
 &=  \frac{\Gamma(3/2)}{\Gamma(1)\Gamma(1/2)} 
 \int_0^1 dx \frac{x^{1/2}(1-x)^{-1/2}}
 {\{k'^2-2x{\bf k}\cdot {\bf k}'+ S(x)\}^{3/2}}
\end{align}
where  $S(x)=x\alpha^2 + xk^2 +(1-x)m^2$. 
We now use the Schwinger's parameterization:
\begin{equation}\label{aeq.2}
\frac{1}{h^\alpha} 
= 
\frac{1}{\Gamma(\alpha)} 
\int^\infty_0 d\tau ~\tau^{\alpha-1} e^{-\tau h},
\end{equation}
Using the above trick to convert the integrand into a Gaussian functional form: 
\begin{align}
 \frac{1}
 {\{k'^2-2x{\bf k}\cdot {\bf k}'+ S(x)\}^{3/2}}
 = \frac{1}{\Gamma(3/2)}\int^\infty_0 d\tau \tau^{1/2} e^{-\tau [k'^2-2x{\bf k}\cdot {\bf k}'+ S(x)] }
\end{align}
Therefore, we have 
\begin{align}
\int\frac{d^D {\bf k}'}{(2\pi)^D} \frac{1}{\alpha^2 + |{\bf k-k'}|^2 }  \frac{1}{( |{\bf k}'|^2 +m^2)^{1/2}}
 &= \frac{1}{\Gamma(1)\Gamma(1/2)} 
 \int_0^1 dx x^{1/2}(1-x)^{-1/2}\int^\infty_0 d\tau \tau^{1/2} e^{-\tau S(x)}
 \nonumber\\& \times 
\int \frac{d^D {\bf k}'}{(2\pi)^D} e^{-\tau [k'^2-2x{\bf k}\cdot {\bf k}']}
\end{align}
 It is also good to remind Gaussian integral formula:
\begin{equation}\label{aeq.6}
\int \frac{d^D {\bf q}}{(2\pi)^D}
e^{ - \frac{1}{2}{\bf q}\cdot \overset\leftrightarrow{\mathbf{A}} \cdot {\bf q}
	+
	 \mathbf{J} \cdot {\bf q}}
= 
\frac{ e^{ \frac{1}{2} \mathbf{J}\cdot \overset\leftrightarrow{\mathbf{A}}^{-1} \cdot \mathbf{J} }}{\sqrt{(2\pi)^D {\rm det}( \overset\leftrightarrow{\mathbf{A}})}},
\end{equation}
where $\overset{\leftrightarrow}{\mathbf{A}}$, $\mathbf{J}$, and $\mathbf{q}$ are defined in a $D$-dimensional space which are 
\begin{align}
     {\bf J} = 2x\tau {\bf k}~~,~~\overset\leftrightarrow{\mathbf{A}} = 2\tau I_D,
\end{align}
with $I_D$ is an identify matrix $D$ dimensions, in the present case. Therefore, we find
\begin{align}
    \int \frac{d^D{\bf k}'}{(2\pi)^D} e^{-\tau [k'^2-2x{\bf k}\cdot {\bf k}']}
    = \frac{e^{\tau x^2 k^2}}{\sqrt{(2\pi)^D(2\tau)^D}}
\end{align}
Accordingly, we can write 
\begin{align}
\int\frac{d^D {\bf k}'}{(2\pi)^D} \frac{1}{\alpha^2 + |{\bf k-k'}|^2 }  \frac{1}{( |{\bf k}'|^2 +m^2)^{1/2}}
 &=  \frac{1}{2^D\pi^{D/2}} \frac{1}{\Gamma(1)\Gamma(1/2)} 
 \int_0^1 dx x^{1/2}(1-x)^{-1/2}
 \nonumber\\&\times
 \int^\infty_0 d\tau \tau^{(1-D)/2} e^{-\tau [S(x)-x^2 k^2]}
\end{align}
Note that 
\begin{align}
 \int^\infty_0 d\tau \tau^{(1-D)/2} e^{-\tau [S(x)-x^2 k^2]} = 
 \frac{\Gamma((3-D)/2)}{[S(x)-x^2 k^2]^{(3-D)/2}}
\end{align}
Therefore, we find 
\begin{align}
\int\frac{d^D {\bf k}'}{(2\pi)^D} \frac{1}{\alpha^2 + |{\bf k-k'}|^2 }  \frac{1}{( |{\bf k}'|^2 +m^2)^{1/2}}
 &=  \frac{\Gamma((3-D)/2)}{2^D\pi^{D/2}} \frac{1}{\Gamma(1)\Gamma(1/2)} 
 \nonumber\\& \times
 \int_0^1 dx 
 \frac{x^{1/2}(1-x)^{-1/2}}{[x\alpha^2 + x(1-x)k^2 +(1-x)m^2]^{(3-D)/2}}
\end{align}
The above integral is regular in two dimensions and therefore, we can safely set $D=2$ which gives
\begin{align}
I_1=\int\frac{d^2 {\bf k}'}{(2\pi)^2} \frac{1}{\alpha^2 + |{\bf k-k'}|^2 }  \frac{1}{( |{\bf k}'|^2 +m^2)^{1/2}}
 =  \frac{1}{4\pi}  
 \int_0^1 dx 
 \frac{x^{1/2}(1-x)^{-1/2}}{[x\alpha^2 + x(1-x)k^2 +(1-x)m^2]^{1/2}}
\end{align}
Similarly, we find 
\begin{align}
    \int \frac{d^D{\bf k}'}{(2\pi)^D} k'_i e^{-\tau [k'^2-2x{\bf k}\cdot {\bf k}'] }
&= 
\left(\frac{1}{2x \tau} \frac{\partial}{\partial k_i} \right) \int \frac{d^D{\bf k}'}{(2\pi)^D} e^{-\tau [k'^2-2x{\bf k}\cdot {\bf k}'] }
\nonumber\\ &= \left(\frac{1}{2x \tau} \frac{\partial}{\partial k_i} \right)\frac{e^{\tau x^2 k^2}}{\sqrt{(2\pi)^D(2\tau)^D}}
= (xk_i) \frac{e^{\tau x^2 k^2}}{\sqrt{(2\pi)^D(2\tau)^D}}
\end{align}
So that, 
\begin{align}
\int\frac{d^D {\bf k}'}{(2\pi)^D} \frac{1}{\alpha^2 + |{\bf k-k'}|^2 }  \frac{k'_i}{( |{\bf k}'|^2 +m^2)^{1/2}}
 &=  \frac{1}{2^D\pi^{D/2}} \frac{1}{\Gamma(1)\Gamma(1/2)} 
 \int_0^1 dx x^{1/2}(1-x)^{-1/2}
 \nonumber\\&\times
 \int^\infty_0 d\tau \tau^{(1-D)/2} (xk_i) e^{-\tau [S(x)-x^2 k^2]}
\end{align}
Thus,  
\begin{align}
\int\frac{d^D {\bf k}'}{(2\pi)^D} \frac{1}{\alpha^2 + |{\bf k-k'}|^2 }  \frac{k'_i}{( |{\bf k}'|^2 +m^2)^{1/2}}
 &=  \frac{k_i}{2^D\pi^{D/2}} \frac{1}{\Gamma(1)\Gamma(1/2)} 
 \int_0^1 dx x^{3/2}(1-x)^{-1/2}
 \nonumber\\&\times
 \int^\infty_0 d\tau \tau^{(1-D)/2} e^{-\tau [S(x)-x^2 k^2]}
\end{align}
Therefore, 
\begin{align}
\int\frac{d^D {\bf k}'}{(2\pi)^D} \frac{1}{\alpha^2 + |{\bf k-k'}|^2 }  \frac{k'_i}{( |{\bf k}'|^2 +m^2)^{1/2}}
 &=  \frac{k_i}{2^D\pi^{D/2}} \frac{\Gamma((3-D)/2)}{\Gamma(1)\Gamma(1/2)} 
 \int_0^1 dx \frac{x^{3/2}(1-x)^{-1/2}}{[S(x)-x^2 k^2]^{(3-D)/2}}
\end{align}
In two dimensions we find 
\begin{align}
I_{2,i}=\int\frac{d^2 {\bf k}'}{(2\pi)^2} \frac{1}{\alpha^2 + |{\bf k-k'}|^2 }  \frac{k'_i}{( |{\bf k}'|^2 +m^2)^{1/2}}
 =  \frac{k_i}{4\pi} 
 \int_0^1 dx \frac{x^{3/2}(1-x)^{-1/2}}{[x\alpha^2 + x(1-x)k^2 +(1-x)m^2]^{1/2}}
\end{align}

Finally, we find 
\begin{align}
&I_1=\int\frac{d^2 {\bf k}'}{(2\pi)^2} \frac{1}{\alpha^2 + |{\bf k-k'}|^2 }  \frac{1}{( |{\bf k}'|^2 +m^2)^{1/2}}
= f_1(k,\alpha,m) 
\label{eq:f1}
\\
&I_2=\int\frac{d^2 {\bf k}'}{(2\pi)^2} \frac{1}{\alpha^2 + |{\bf k-k'}|^2 }  \frac{\hat\sigma\cdot{\bf k}'}{( |{\bf k}'|^2 +m^2)^{1/2}}
 =  (\hat\sigma\cdot{\bf k}) f_3(k,\alpha,m) 
 \label{eq:f3}
\end{align}
where we define function $f_n(k,\alpha,m)$: 
\begin{align}
 f_n(k,\alpha,m)= \frac{1}{4\pi}\int_0^1 dx \frac{x^{n/2}(1-x)^{-1/2}}{[x\alpha^2 + x(1-x)k^2 +(1-x)m^2]^{1/2}}
 = a_n + O(k^2)
\end{align}
Now, we perform the integration over $x$ for small $k$ limit: 
\begin{align}
&a_1 = \frac{m}{2\pi} \frac{K\left(1-\frac{\alpha^2 }{m^2}\right)-E\left(1-\frac{\alpha^2
   }{m^2}\right)}{m^2-\alpha^2 }
\\
&a_3= \frac{m}{6\pi}\frac{ \left(3 m^2-\alpha^2 \right)
   K\left(1-\frac{\alpha^2 }{m^2}\right)+2  \left(\alpha^2 -2 m^2\right)
   E\left(1-\frac{\alpha^2 }{m^2}\right)}{(m^2-\alpha^2)^2}
\end{align}
We find the following asymptotic behaviors 
\begin{align}
& a_1 \approx \frac{1}{2\pi} \frac{1}{\alpha}, \hspace{19mm}\text{for $\alpha\gg m$ }
\\
& a_1 \approx  \frac{1}{2\pi}\frac{\ln(m/\alpha)}{m}, \hspace{9mm} \text{for $\alpha\ll m$} \label{eq:small-alpha}
\end{align}
and
\begin{align}
&a_3 \approx \frac{1}{3\pi}\frac{1}{\alpha}, \hspace{19mm}\text{for $\alpha\gg m$ }
\\
& a_3 \approx \frac{1}{2\pi} \frac{\ln(m/\alpha)}{m}. \hspace{9mm} \text{for $\alpha\ll m$}
\end{align}
We conclude this section by expressing the self-energy part ${\bm\Xi}$, as given in Eq. \eqref{self-Xi}, through the substitution of Eqs. \eqref{eq:f1} and \eqref{eq:f3}. The resulting form is presented as
\begin{align}
    {\bm\Xi} = {\revise{\frac{g^2_{\rm em}}{2D (\hbar v_F k_\circ) }}} \left\{ - f_3(k,\alpha,m) \hbar v_F {\bf k}\cdot{\bm \sigma} - f_1(k,\alpha,m) m_z\sigma_z , \label{self-Xi-final}\right\}
\end{align} 
which agrees with the form presented in the main text as Eq. (13). 

\begin{figure}[t]
\centering
\includegraphics[width=0.99\textwidth]{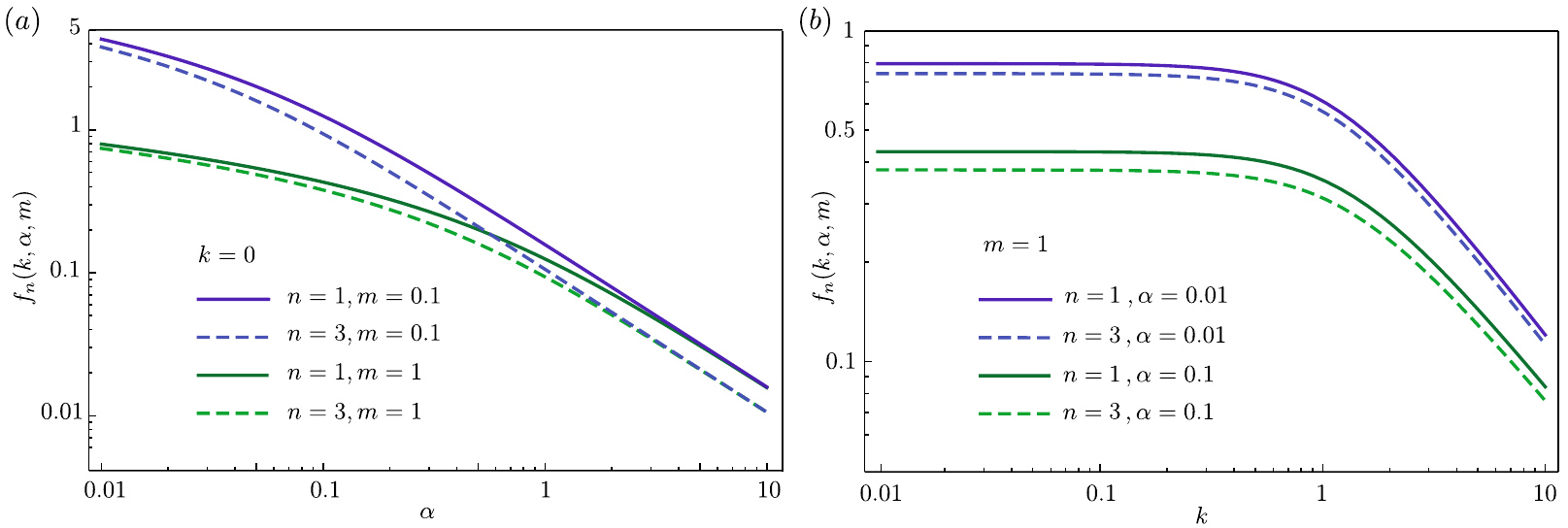}
\caption{$f_1$ and $f_3$ functions in logarithmic scales versus (a) $\alpha$ at $k=0$, and (b) $k$ for $m=1$. For comparison, two different values for $m$ and $\alpha$ have been considered in panels (a) and (b), respectively.}
\end{figure}

\subsection{Self-energy and renormalized energy dispersion of surface states}
\begin{figure}[t]
\centering
\includegraphics[width=0.5\textwidth]{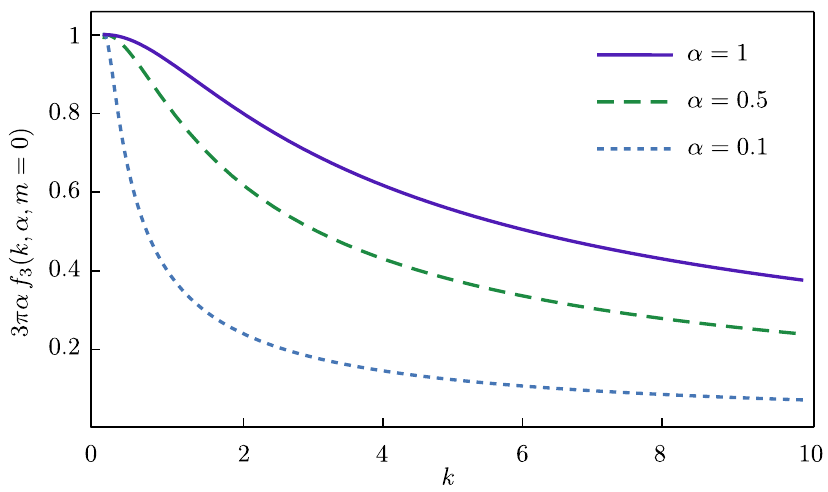}
\caption{Momentum dependence of $f_3(k,\alpha,0)$}
\label{fig:f3-m0}
\end{figure}

Here, we elaborate the results for the self-energy 
${\bm \Sigma}$ in a bit more detail.  
Let's first concentrate on the case of side surfaces parallel to the magnetization which has gapless surface states. 
For these surfaces, we see that the influence of self-energy consists of different forms of velocity renormalizations and a tilting term
depending on the detailed matrix structure of magnon-electron coupling.
Considering four interesting situations of completely isotropic Heisenberg, pure DM term, Kitaev type and highly anisotropic coupling to the magnons, which correspond to ${\bf n}_{H}=\hat{\bf x}+i\hat{\bf y}$, ${\bf n}_{DM}=\hat{\bf y}-i\hat{\bf x}$,
${\bf n}_{K}=\hat{\bf y}+i\hat{\bf x}$,
and ${\bf n}_{\rm I}=\hat{\bf y}$, respectively,
we can easily see that
\begin{align}
     &     
     {\bm \Sigma}_{H} = {\bm \sigma}.\hat{\bf n}_{H}\:{\bm \Xi} \:{\bm \sigma}.\hat{\bf n}_{H}^\ast = 
     -(\sigma_z+\sigma_0) \Xi_{z},
     \\
     &
     {\bm \Sigma}_{DM} = {\bm \sigma}.\hat{\bf n}_{DM}\:{\bm \Xi} \:{\bm \sigma}.\hat{\bf n}_{DM}^\ast = 
     -(\sigma_z+\sigma_0) \Xi_{z}, 
     \\
     &
     {\bm \Sigma}_{K} = {\bm \sigma}.\hat{\bf n}_{K}\:{\bm \Xi} \:{\bm \sigma}.\hat{\bf n}_{K}^\ast = 
     -(\sigma_z-\sigma_0) \Xi_{z}, 
     \\
     &
    {\bm \Sigma}_{I}|_{\hat{\bf n}=\hat{\bf y}} = {\sigma}_y\:{\bm \Xi} \:{\sigma}_y = 
     -\sigma_x \Xi_{x}
     -\sigma_z \Xi_{z}. 
\end{align}
Therefore, for the side surfaces elongated in $xz$ plane, the self-energy corrections are   
\begin{align}
   & {\bm \Sigma}_{\rm H,DM}(k_x,k_z) =  
   {\revise{\frac{g^2_{\rm em}}{2D (\hbar v_F k_\circ) }}}
   f_3( \sqrt{k_x^2+k_z^2},\alpha,0) \: \hbar v_F k_z \big(\sigma_z+\sigma_0\big), \\
      & {\bm \Sigma}_{\rm K}(k_x,k_z) = 
      {\revise{\frac{g^2_{\rm em}}{2D (\hbar v_F k_\circ) }}}
      f_3( \sqrt{k_x^2+k_z^2},\alpha,0) \: \hbar v_F k_z \big(\sigma_z-\sigma_0\big), \\
   &
    {\bm \Sigma}_{\rm I}(k_x,k_z) =  
    {\revise{\frac{g^2_{\rm em}}{2D (\hbar v_F k_\circ) }}}
    f_3( \sqrt{k_x^2+k_z^2},\alpha,0) \:  \hbar v_F \big( k_x\sigma_x + k_z \sigma_z\big).
\end{align}
In fact, depending on the form of magnon-electron coupling terms,
we can have anisotropy besides a Dirac cone tilting, and just an overall isotropic velocity renormalization in the two cases, respectively. The effect of self-energies can be constituted to a dimensinsless renormalization factor $f_3(k,\alpha,0)$ which has a strong momentum dependence and vanishes
in a power-law form for large enough momenta. 
This can be clearly seen by writing down the explicit form of 
$f_3(k,\alpha,0)$ which reads
\begin{align}
f_3(k,\alpha,0) = 
-\frac{\sqrt{\alpha ^2+k^2}}{4 \pi  k^2}
+\frac{\left(\alpha ^2+2 k^2\right)}{4 \pi  k^3}\,\ln \left(\frac{\sqrt{\alpha ^2+k^2}+k}{\alpha }\right)
\end{align}
As this expression and its illustration in Fig. \ref{fig:f3-m0} show,
the renormalization becomes negligible for $k \gg \alpha$. 
In addition, its amplitude at very low momenta scales inversely with $\alpha$
as it is followed from the limiting form of the function $f_3$ which is given by
\begin{align}
    f_3(k,\alpha,0) \approx   \frac{1}{3 \pi \alpha }-\frac{ k^2}{30\pi \alpha ^3}+O\big(k^4\big).
\end{align}
An immediate consequence of this result is that when the magnon gap becomes very small ($\alpha\to 0$), we find a very radical change in the effective velocity of electrons but only in the vicinity of very small $k\lesssim \alpha$ as it has been illustrated in Fig. 2 (b) and (c) in the main text.

Now, we return to the top surface where we have a magnetic gap $m_z$
and considering isotropic coupling to the magnons (${\bf n}_{\rm I}=\hat{\bf x}+i\hat{\bf y}$), the matrix structure of interaction only leaves the term proportional $m_z$ and we have
\begin{align}
    {\bm \Sigma}_{\rm top}(k_x,k_y) &= 
    {\revise{\frac{g^2_{\rm em}}{2D (\hbar v_F k_\circ) }}}
    m_z\: f_1(\sqrt{k_x^2+k_y^2},\alpha,m_z) \: \big(\sigma_z+\sigma_0\big).
\end{align}
where 
\begin{align}
    f_1(k,\alpha,m_z) = \frac{1}{2\pi k}\ln \left(\frac{\sqrt{\alpha ^2+k^2}+k}{\alpha }\right)+ O(m^2_z) \approx \frac{1}{2\pi \alpha}-\frac{k^2}{12\pi \alpha^3} + O(k^4).
\end{align}

When the magnon gap is large enough such that ($\alpha>m_z\neq 0$), the self-energy will modify the band gap by $m_z/(2\pi\alpha)$, whereas in the opposite limit of vanishingly small magnon gap for which $\alpha\ll m_z$, we find logarithmic correction term $ (1/2\pi)\ln(m_z/\alpha)$
as can be seen from Eq. \eqref{eq:small-alpha}.

\section{Vanishing Two-loop self-energy for an infinite range uniform interaction}
\begin{figure}[b]
    \centering
    \includegraphics[width=0.7\linewidth]{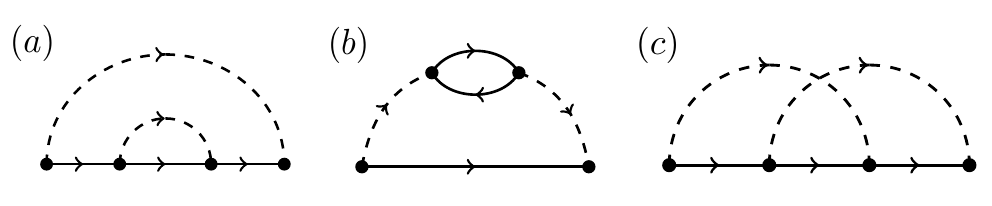}
    \caption{Feynman diagrams for the two-loop self-energy. Solid lines represent the electron propagator, while dashed lines indicate the interaction potential.}
    \label{fig2loop}
\end{figure}

For ultra-long-range interaction, the first-order self-energy is the dominant contribution. In this section, we prove that second-order self-energy vanishes for an infinite-range uniform interaction , i.e. $V({\bm q})=V\delta({\bm q})$.
The second-order self-energy can be decomposed in three parts as depicted in Feynman diagrams shown in Fig.~\ref{fig2loop}. Therefore, we can write 
\begin{align}
\hat \Sigma^{(2)}({\bm k},i\omega) =\hat \Sigma^{(2a)}({\bm k},i\omega) + \hat \Sigma^{(2b)}({\bm k},i\omega) + \hat \Sigma^{(2c)}({\bm k},i\omega) 
\end{align}
in which, we have
\begin{align}
\hat \Sigma^{(2a)}({\bm k},i\omega) &=i \int \frac{d^D {\bm q}}{(2\pi)^D}  \int^{\infty}_{-\infty} \frac{dE}{2\pi} 
V({\bm q}) \hat G(i(\omega-E),{\bm k}-{\bm q}) 
\hat \Sigma^{(1)}({\bm k}-{\bm q}) \hat G(i(\omega-E),{\bm k}-{\bm q})~,
\\
\hat \Sigma^{(2b)}({\bm k},i\omega) &=i \int \frac{d^D {\bm q}}{(2\pi)^D}  \int^{\infty}_{-\infty} \frac{dE}{2\pi} 
V({\bm q})^2 \chi({\bm q},iE) 
\hat G(i(\omega-E),{\bm k}-{\bm q})~,
 \\
\hat \Sigma^{(2c)}({\bm k},i\omega) &=i^2 \int \frac{d^D {\bm q}}{(2\pi)^D} \int \frac{d^D {\bm q}'}{(2\pi)^D} 
 \int^{\infty}_{-\infty} \frac{dE}{2\pi} \int^{\infty}_{-\infty} \frac{dE'}{2\pi} 
 V({\bm q}) V({\bm q}')\hat G(i(\omega-E),{\bm k}-{\bm q})  
\nonumber\\&
\hat G(i(\omega-E-E'),{\bm k}-{\bm q}-{\bm q}')  
 \hat G(i(\omega-E'),{\bm k}-{\bm q}')  
\end{align}
where $\hat \Sigma^{(1)}({\bm k})$ stands for the first order self-energy and the bare susceptibility, $\chi({\bm q},E)$, reads 
\begin{align}
\chi({\bm q},i\omega) &=-i  \int \frac{d^D {\bm k}}{(2\pi)^D} 
\int^{\infty}_{-\infty} \frac{dE}{2\pi} {\rm Tr} \Big[  \hat G(i(E+\omega),{\bm k}+{\bm q})
 \hat G(iE,{\bm k}) \Big ]
\end{align}
Here, we restrict ourselves infinite range uniform interaction which implies $V({\bm q}) = V \delta({\bm q})$ and thus $\hat \Sigma^{(1)}({\bm k}) = V \hat{\bm \sigma}\cdot \hat {\bm k}/2$ \footnote{H, Rostami, E. Cappelluti, and A. V. Balatsky,
Phys. Rev. B 98, 245114, (2018)}. 
Therefore, we find
\begin{align}
\hat \Sigma^{(2a)}({\bm k},i\omega) &=i \frac{V^2}{2(2\pi)^D} \int^{\infty}_{-\infty} \frac{dE}{2\pi} 
\hat G(i(\omega-E),{\bm k}) \hat{\bm \sigma}\cdot\hat{\bm k} 
\hat G(i(\omega-E),{\bm k}) 
 \\
 \hat \Sigma^{(2b)}({\bm k},i\omega) &=i \frac{V^2 }{(2\pi)^D}    \int^{\infty}_{-\infty} \frac{dE}{2\pi} 
 [\lim_{\bm q \to \bm 0} \delta({\bm q})\chi({\bm q},iE) ]
 \hat G(i(\omega-E),{\bm k}) 
 \\
\hat \Sigma^{(2c)}({\bm k},i\omega) &=i^2  \frac{V^2}{(2\pi)^{2D}}
 \int^{\infty}_{-\infty} \frac{dE}{2\pi} \int^{\infty}_{-\infty} \frac{dE'}{2\pi} 
 \hat G(i(\omega-E),{\bm k})  
 \hat G(i(\omega-E-E'),{\bm k})   \hat G(i(\omega-E'),{\bm k})  
\end{align}
In the following, we calculate each of these self-energies separately. For the case of Fig.~2(a), we have  ($\hbar v=1$)
\begin{align}
&\hat \Sigma^{(2a)}({\bm k},i\omega) =i \frac{V^2}{2(2\pi)^D} \int^{\infty}_{-\infty} \frac{dE}{2\pi} 
\hat G(i(\omega-E),{\bm k}) \hat{\bm \sigma}\cdot\hat{\bm k}
\hat G(i(\omega-E),{\bm k}) 
\nonumber\\
&= i \frac{V^2}{2(2\pi)^D}
 \int^{\infty}_{-\infty} \frac{dE}{2\pi} 
\frac{i(\omega-E)+{\bm \sigma}\cdot{\bm k}}{(E-\omega)^2+k^2} 
\hat{\bm \sigma}\cdot\hat{\bm k} \frac{i(\omega-E)+{\bm \sigma}\cdot{\bm k}}{(E-\omega)^2+k^2} 
\end{align}
where we must evaluate the following integral 
\begin{align}
\int^{\infty}_{-\infty} \frac{dE}{2\pi} 
\frac{[i(\omega-E)+k\hat{\bm \sigma}\cdot\hat{\bm k}] \hat{\bm \sigma}\cdot\hat{\bm k} [i(\omega-E)+k\hat{\bm \sigma}\cdot\hat{\bm k}] }
{[(E-\omega)^2+k^2]^2} 
\end{align}
Note that we have
\begin{align}
[i(\omega-E) \mathbbm{I}+k\hat{\bm \sigma}\cdot\hat{\bm k}] \hat{\bm \sigma}\cdot\hat{\bm k} [i(\omega-E) \mathbbm{I}+k\hat{\bm \sigma}\cdot\hat{\bm k}]
= 
[k^2-(E-\omega)^2] \hat{\bm \sigma}\cdot\hat{\bm k} - 2i k(E-\omega)  \mathbbm{I}
\end{align}
which implies
\begin{align}
&\int^{\infty}_{-\infty} \frac{dE}{2\pi} \frac{ [k^2-(E-\omega)^2] \hat{\bm \sigma}\cdot\hat{\bm k} - 2i k(E-\omega)  \mathbbm{I}  }{[(E-\omega)^2+k^2]^2} 
=
\int^{\infty}_{-\infty} \frac{dE}{2\pi} \frac{ [k^2-E^2] \hat{\bm \sigma}\cdot\hat{\bm k} - 2i k E  \mathbbm{I} }{[E^2+k^2]^2} 
\nonumber\\&=
\hat{\bm \sigma}\cdot\hat{\bm k} \int^{\infty}_{-\infty} \frac{dE}{2\pi} \frac{k^2-E^2}{(E+ik)^2(E-ik)^2} =0
\end{align}
Eventually, we find 
$
\hat \Sigma^{(2a)}({\bm k},i\omega) =0. 
$
For the case of Fig.~2(b), we first need to evaluate the susceptibility which follows 
\begin{align}
\chi({\bm 0},i\omega) &=-i  \int \frac{d^D {\bm k}}{(2\pi)^D} 
\int^{\infty}_{-\infty} \frac{dE}{2\pi} {\rm Tr} \Big[  
\frac{i(E+\omega)+{\bm \sigma}\cdot{\bm k}}{(E+\omega)^2+k^2} 
\frac{i E+{\bm \sigma}\cdot{\bm k}}{E^2+k^2} 
\Big ]
\end{align}
Note that 
\begin{align}
&{\rm Tr} \left[  
\frac{i(E+\omega) \mathbbm{I}+{\bm \sigma}\cdot{\bm k}}{(E+\omega)^2+k^2} 
\frac{i E  \mathbbm{I} +{\bm \sigma}\cdot{\bm k}}{E^2+k^2} 
\right ] 
=
D \frac{k^2-E(E+\omega)}{(E^2+k^2)((E+\omega)^2+k^2)}
\end{align}
Therefore, 
\begin{align}
\chi({\bm 0},i\omega) =-i  D  \int \frac{d^D {\bm k}}{(2\pi)^D} 
\int^{\infty}_{-\infty} \frac{dE}{2\pi}  \frac{k^2-E(E+\omega)}{(E^2+k^2)((E+\omega)^2+k^2)}
\end{align}
It can be shown that the energy integration vanishes: 
\begin{align}
\int^{\infty}_{-\infty} \frac{dE}{2\pi}  \frac{k^2-E(E+\omega)}{(E^2+k^2)((E+\omega)^2+k^2)} =0
\end{align}
Implying that $\chi({\bm 0},i\omega) =0$. Vanishing susceptibility at $q=0$ is not a surprise and it is granted by gauge invariance and charge conservation that lead~\footnote{G. F. Giuliani and G. Vignale, {\em Quantum Theory of the Electron Liquid}, (Cambridge University Press,Cambridge, 2005).}
\begin{align}
\chi({\bm q},\omega) = \left(\frac{q}{\omega} \right)^2 \chi_{J_L J_L}({\bm q},\omega)
\end{align} 
where $\chi_{J_L J_L}({\bm q},\omega)$ stands for the longitudinal part of current-current correlation function. 
Although $\lim_{{\bm q }\to \bm 0} \delta({\bm q}) \to \infty$, the self-energy (2b), $\hat \Sigma^{(2b)}({\bm k},i\omega)$, is proportional to 
$\lim_{{\bm q }\to \bm 0} \chi({\bm q},i\omega) \delta({\bm q}) \to 0$.
Therefore, we find 
$
\hat \Sigma^{(2b)}({\bm k},i\omega) =0. 
$
The last part of self-energy reads
\begin{align}
\hat \Sigma^{(2c)}({\bm k},i\omega) &=i^2  \frac{V^2}{(2\pi)^{2D}}
 \int^{\infty}_{-\infty} \frac{dE}{2\pi} \int^{\infty}_{-\infty} \frac{dE'}{2\pi} 
 \hat G(i(\omega-E),{\bm k})  
 \hat G(i(\omega-E-E'),{\bm k})   \hat G(i(\omega-E'),{\bm k})  
\end{align}
For which following double integration is required to be performed. 
\begin{align}
&\int^{\infty}_{-\infty} \frac{dE}{2\pi} \int^{\infty}_{-\infty} \frac{dE'}{2\pi} 
\frac{i(\omega-E) \mathbbm{I}+  \hat {\bm \sigma}\cdot {\bm k}}{(\omega-E)^2+k^2} 
\frac{i(\omega-E-E') \mathbbm{I}+  \hat {\bm \sigma}\cdot {\bm k}}{(\omega-E-E')^2+k^2} 
\frac{i(\omega-E') \mathbbm{I}+  \hat {\bm \sigma}\cdot {\bm k}}{(\omega-E')^2+k^2} 
\end{align}
Note that
\begin{align}
&[i(\omega-E) \mathbbm{I}+  \hat {\bm \sigma}\cdot {\bm k}]
[i(\omega-E-E') \mathbbm{I}+  \hat {\bm \sigma}\cdot {\bm k}]
[i(\omega-E') \mathbbm{I}+  \hat {\bm \sigma}\cdot {\bm k}]
\nonumber \\&
= i A(k,\omega,E,E')  \mathbbm{I}+ B(k,\omega,E,E') \hat{\bm\sigma}\cdot {\bm k}
\end{align}
in which
\begin{align}
A(k,\omega,E,E') &= (\omega-E)  [k^2 -(\omega-E-E') (\omega-E')]
+ [2(\omega-E')-E]k^2~, 
\\
B(k,\omega,E,E') &=k^2 -(\omega-E-E') (\omega-E') 
-(\omega-E)  [2(\omega-E')-E]~.
\end{align}
One can prove that 
\begin{align}
\int^{\infty}_{-\infty} \frac{dE}{2\pi} \int^{\infty}_{-\infty} \frac{dE'}{2\pi} 
\frac{A(k,\omega,E,E')}{C(k,\omega,E,E')} =0~~~,~~~
\int^{\infty}_{-\infty} \frac{dE}{2\pi} \int^{\infty}_{-\infty} \frac{dE'}{2\pi} 
\frac{B(k,\omega,E,E') }{C(k,\omega,E,E')} =0
\end{align}
where $C(k,\omega,E,E')=[(\omega-E)^2+k^2][(\omega-E-E')^2+k^2][(\omega-E')^2+k^2]$. 
Therefore, we conclude 
$
\hat \Sigma^{(2c)}({\bm k},i\omega) =0. 
$